\documentclass[a4paper,10pt,twocolumn,superscriptaddress,nofootinbib]{revtex4}
\usepackage{amsmath,amssymb, placeins}
\usepackage[final]{graphicx}
\usepackage[english]{babel}
\usepackage[utf8]{inputenc}
\usepackage{hyperref}

\graphicspath{{./}}
\makeatletter
\renewcommand{\p@subsection}{}
\renewcommand{\p@subsubsection}{}
\makeatother

    \def\<{\langle}
    \def\>{\rangle}

  \newcommand{\eq}[1]{
    \begin{equation}
    {#1}
    \end{equation}}

    \newcommand{\nmq}[1]{
    \begin{multline}
    #1
    \end{multline}}

\begin{document}

\title{Fractional quantum Hall effect in bilayer graphene beyond the single Landau level approximation}
\author{Kyrylo Snizhko}
\email{snezhkok@gmail.com}
\affiliation{Physics Department, Taras Shevchenko National University of Kyiv, Kyiv, 03022, Ukraine}
\affiliation{Department of Physics, Lancaster University, Lancaster, LA1~4YB, UK}
\author{Vadim Cheianov}
\affiliation{Department of Physics, Lancaster University, Lancaster, LA1~4YB, UK}
\author{Steven H. Simon}
\affiliation{Rudolf Peierls Centre for Theoretical Physics, Oxford University, Oxford, OX1~3NP, UK}

\begin{abstract}
Bilayer graphene has been predicted to give unprecedented tunability of the electron-electron interaction with the help of external parameters, allowing one to stabilize different fractional quantum Hall states. Recent experimental works make theoretical analysis of such systems extremely relevant. In this paper we describe a methodology for investigating the possibility of realizing specific fractional quantum Hall states in bilayer graphene taking into account polarization effects and virtual interband transitions. We apply this methodology to explore the possibility of realizing the Moore-Read Pfaffian state in bilayer graphene.
\end{abstract}

\maketitle

\tableofcontents
\newpage

\section*{Introduction}

Since the discovery of graphene \cite{GeimNovoselovGraphene} it has been well-understood that this material opens new horizons on the investigation of quantum Hall effects (QHE). The reasons for this are a much better confinement of the electron gas in a 2D plane, longer mean free paths, and larger cyclotron gaps than can be achieved in GaAs quantum wells. The integer and fractional QHE have been observed in systems with suspended graphene and graphene on different substrates \cite{IQHE_BerryPhase_graphene, Bolotin2009, YacobyFQHE_graphene_2012, FQHE_graphene_hBN_obsrevation}. The sequence of conducting plateaux in graphene was found to differ from that in GaAs, which is due to a peculiar structure of the single particle spectrum.

Bilayer graphene (BLG) has attracted attention of the fractional quantum Hall effect (FQHE) community for some time. The main reason for that is that this material allows for unprecedented tunability of the parameters important for FQHE by means of external electric and magnetic fields \cite{Apalkov2011, SniCheSim}. It turns out that the effects of Landau level mixing, vacuum polarization etc. are extremely important in this system \cite{SniCheSim}, making it complicated to analyze theoretically. However, recent experimental advances in observation of FQHE in BLG \cite{Yacobi_Exp_BLG_EHAsymmFQHE, MorpurgoFalko_HalfFilling_BLG_Observation, TunableFQHE_BLG_Observation} call for theoretical analysis. This paper expounds in detail a microscopic study of the possibility of realizing the so-called Moore-Read Pfaffian FQHE state \cite{MooreRead} in BLG, which was briefly reported in \cite{SniCheSim}.

We perform the analysis of Landau level mixing, vacuum polarization and some other effects in order to establish the conditions under which theoretical predictions are reliable. Under appropriate conditions, we treat level mixing pertubatively in the spirit of \cite{Morf_Screening, RezayiHaldane_LLmixing_perturb} and investigate the possibility of realizing the Moore-Read state using exact numerical diagonalization. The methodology presented here can be readily applied to study the possibility of realizing any other FQHE state in BLG given the state's trial wave function.

The first section reminds the reader of the methodology for studying FQHE in conventional ("non-relativistic") systems based on exact numerical diagonalization of the system Hamiltonian in the single Landau level approximation.

In the second section we describe a formalism which allows the use of the same methodology in bilayer graphene.

In the third section we discuss the effects of interaction of different Landau levels and incorporation of them into the single-Landau-level-based methodology introduced earlier.

Finally, in the fourth section we apply our methodology that takes into account the deviations from the single Landau level approximation to study the possibility of realizing the Moore-Read \cite{MooreRead} state in bilayer graphene.

\section{\label{Sect_NumDiagNonRel}Numerical diagonalization approach to the quantum Hall effect in non-relativistic systems}

In this section we recall how Landau levels (LLs) emerge in a two-dimensional non-relativistic system in a magnetic field. We also introduce some notation that will be used in the following sections.

\subsection{Problem of a single non-relativistic electron in a magnetic field}

The single electron Hamiltonian in the uniform magnetic field, perpendicular to the plane of the system, is
\begin{equation}
\label{H_NR}
H_{1-\mathrm{part}} = \frac{\boldsymbol{\pi}^{2}}{2m^*} - g \mu_{\mathrm{B}} B S_{z},
\end{equation}
where $\boldsymbol{\pi} = (\pi_x, \pi_y)$ (since the system is two-dimensional), $\pi_i = p_i + e A_i/c$, $p_i = -i \hbar \partial_i$, $e$ is the elementary charge, $\mathbf{A}=[\mathbf{B}\times\mathbf{r}]/2$ is the vector potential of the uniform magnetic field $\mathbf{B} = -B \mathbf{e}_z$, $S_z$ is the $z$-component of the electron spin, $m_e$ is the free electron mass, $m^*$ is the effective mass of an electron, $g$ is the Lande $g$-factor, and $\mu_{\mathrm{B}} = e \hbar/(2 m_e c)$ is the Bohr magneton.

We introduce the magnetic length $l$, the cyclotron frequency $\omega_c$, and the complex coordinate $w$ in the plane:
\begin{equation}
l = \sqrt{\frac{\hbar c}{e B}}\ ,\ \omega_c = \frac{e B}{m^* c}\ ,\ w = \frac{x+i y}{l},\ \bar{w} = \frac{x-i y}{l}.
\end{equation}

We also introduce operators $\hat{a}, \hat{a}^\dag, \hat{b}, \hat{b}^\dag$:
\begin{eqnarray}
\label{ab-operatorsb}
\hat{a} = \sqrt{2}\left(\bar{\partial}+\frac{w}{4}\right),\quad \hat{a}^\dag = \sqrt{2}\left(-\partial+\frac{\bar{w}}{4}\right),\\
\hat{b} = \sqrt{2}\left(\partial+\frac{\bar{w}}{4}\right),\quad \hat{b}^\dag = \sqrt{2}\left(-\bar{\partial}+\frac{w}{4}\right).
\label{ab-operatorse}
\end{eqnarray}
where $\partial$ and $\bar{\partial}$ denote $\partial/\partial w$ and $\partial/\partial \bar{w}$ respectively. All commutation relations between these four operators are trivial except for the following:
\begin{equation}
\left[\hat{a}, \hat{a}^\dag\right] = \left[\hat{b}, \hat{b}^\dag\right] = 1.
\end{equation}

We can then rewrite the Hamiltonian in the form
\begin{equation}
H_{1-\mathrm{part}} = \hbar \omega_c \left(\hat{a}^\dag \hat{a} + \frac{1}{2}\right) - g \mu_{\mathrm{B}} B S_{z}.
\end{equation}

The operators $\hat{a}, \hat{a}^\dag$ are similar to the ladder operators in the problem of the harmonic oscillator, thus the system's spectrum consists of Landau levels with energies $E_n = \hbar \omega_c (n+1/2) - g \mu_{\mathrm{B}} B S_{z}$, $n \in \mathbb{Z}_{+}$, $S_{z} = \pm 1/2$. Operators $\hat{b}, \hat{b}^\dag$ commute with the Hamiltonian, thus they transform one state into another state within the same Landau level.

Let us consider the operator of the $z$-projection of the orbital angular momentum:
\begin{equation}
\label{Ldef}
\hat{L} = \hat{L}_z / \hbar = \left[\mathbf{r} \times \mathbf{p}\right]_z / \hbar = z \partial - \bar{z} \bar{\partial} = \hat{b}^\dag \hat{b} - \hat{a}^\dag \hat{a}.
\end{equation}

One can see that
\begin{gather}
\left[\hat{L},\hat{a}\right]=\hat{a}, \left[\hat{L},\hat{a}^\dag\right]=-\hat{a}^\dag, \left[\hat{L},\hat{b}\right]=-\hat{b}, \left[\hat{L},\hat{b}^\dag\right]=\hat{b}^\dag\\
\Rightarrow \left[\hat{L},H_{1-\mathrm{part}} \right]=0.
\end{gather}

Thus the eigenstates of the Hamiltonian \eqref{H_NR} can be labeled by three quantum numbers: the Landau level number $n \in \mathbb{Z}_+$, the $z$-projection of the angular momentum $m~\in~(\mathbb{Z}_+ - n)$ and the electron spin projection \mbox{$S_z = \pm 1/2$}
\begin{equation}
|n,m,S_z\rangle = \psi_{nm}(w) \otimes |S_z\rangle.
\end{equation}

The orbital wave function $\psi_{nm}$ can be expressed as
\begin{eqnarray}
\label{NRwfb}
\psi_{nm}(w) &=& \frac{1}{\sqrt{n! (n+m)!}} (\hat{b}^\dag)^{n+m} (\hat{a}^\dag)^{n}\psi_{00}(w),\\
\psi_{00}(w) &=& \frac{1}{\sqrt{2 \pi l^2}} e^{-\frac{|w|^2}{4}},\quad \hat{a}\psi_{00} = \hat{b}\psi_{00} = 0.
\label{NRwfe}
\end{eqnarray}
Note that the wave functions \eqref{NRwfb} are polynomials of complex coordinates $w, \bar{w}$, multiplied by the exponential $\exp{(-|w|^2/4)}$ which is the same for all of the states. In particular,
\begin{equation}
\label{zerothLLwf}
\psi_{n=0,m}(w) = \frac{1}{\sqrt{2^{m+1} \pi l^2}} w^m e^{-\frac{|w|^2}{4}}.
\end{equation}

So, the system's energy levels are Landau levels with energies $E_{n,S_z} = \hbar \omega_c (n+1/2) - g \mu_{\mathrm{B}} B S_{z}$, with eigenstates in a LL labeled by the angular momentum projection $m \geq -n$.

In a finite sample there is only a finite number of states available to an electron in a Landau level. One can estimate their number using the fact that the states \eqref{NRwfb}-\eqref{NRwfe} are spatially localized: the number of states in a Landau level of a finite round sample is approximately equal to the number of states \eqref{NRwfb}-\eqref{NRwfe} that are localized mainly in the area of the sample. Thus, one can introduce the filling factor $\nu$:
\begin{equation}
\nu = 2 \pi l^2 n_{e} = N_{e}/N_{\text{orb}},
\end{equation}
where $n_{e}$ is the density of electrons, $N_{e}$ is the total number of electrons in the system and $N_{\mathrm{orb}}$ is the number of orbitals in a LL in the sample.

Typically in GaAs heterostructures $m^{*} \approx 0.07 m_e$, $g \approx -0.4$, thus LLs with the same number $n$ but different spin projections form closely spaced doublets. For a typical fractional quantum Hall (FQH) experiment in such systems the characteristic Coulomb energy scale $e^2/(\kappa l)$ ($\kappa$ is the dielectric constant, in GaAs $\kappa \approx 13$) is on the same order as the cyclotron frequency $\hbar \omega_c$. For $B \gtrsim 5\text{T}$ the interaction energy scale is less than the inter-doublet spacing. Therefore, in these conditions the Coulomb interaction cannot throw the electrons from a Landau level to other Landau levels efficiently.

In this case for any filling factor one can expect only one doublet to be partially filled with others either fully filled or completely empty. Thence, it is not too bad an approximation to restrict consideration to the electrons in the partially filled doublet. Corrections to this picture can be taken into account by means of perturbation theory \cite{Morf_Screening, RezayiHaldane_LLmixing_perturb, SimonRezayi_LLmixing_perturb, PetersonNayak_LLmixing_perturb, SodemannMacDonald_LLmixing_perturb, PetersonNayak2_LLmixing_perturb}. However, in this section we neglect them.

In a partially filled doublet, the lowest energy is usually achieved when the electrons form a spin-polarized state. In such a state only one of the two Landau levels in a doublet is partially occupied, and the other is either empty or fully occupied. There are two reasons for that. One is that such states minimize the Coulomb interaction exchange energy (if the interaction potential decreases monotonically with distance, which is typically the case). Another reason for the electrons to form spin-polarized states is the Zeeman splitting (even though it is small).

In the remaining part of this section we only consider one Landau level, neglecting the influence of other Landau levels. This approximation is called the single Landau level approximation (SLLA). We also assume that the state is spin-polarized, therefore we suppress the spin variables.

\subsection{\label{SubSect_TwoElectrProblem_NonRel}Interaction of two electrons in a non-relativistic Landau level}

We begin the discussion of the many-body problem with the two-particle case. For interaction potentials which depend only on the distance between the electrons this problem can be solved exactly. This solution gives an opportunity to introduce some important notions.

The two-electron Hamiltonian can be written as follows:
\begin{eqnarray}
\hat{H}_{2-\mathrm{part}} &=& \hat{H}_{\mathrm{free}}+V(r),\\
\hat{H}_{\mathrm{free}} &=& \hat{H}_{1-\mathrm{part},1}+\hat{H}_{1-\mathrm{part},2},
\end{eqnarray}
where $r = |\mathbf{r}_1-\mathbf{r_2}| = l |w_1-w_2|$, $V(r)$ is the interaction potential, e.g., the Coulomb potential.

Since we are working in the SLLA approximation, the single-particle part of the Hamiltonian is proportional to the identity operator and can be excluded from the consideration. Thus to diagonalize the Hamiltonian we only need to diagonalize the interaction potential operator $V(r)$ in the Hilbert space spanned by vectors
\begin{equation}
|m_1,m_2\rangle = \frac{1}{\sqrt{2}}(|m_1\rangle \otimes |m_2\rangle - |m_2\rangle \otimes |m_1\rangle),
\end{equation}
with the angular momenta of the two electrons, $m_1$ and $m_2$, taking all the possible values in the LL considered.

We introduce $z$-projections of the relative angular momentum and the angular momentum of the center of mass:
\begin{multline}
\hat{L}_{\mathrm{rel}}=\left(\frac{1}{2\hbar}[(\mathbf{r}_1-\mathbf{r_2})\times(\mathbf{p}_1-\mathbf{p}_2)]\right)_z = {}\\
{}\frac{1}{2}\left(\hat{L}_1+\hat{L}_2-b_1^\dag b_2-b_2^\dag b_1+a_1^\dag a_2+a_2^\dag a_1\right),
\end{multline}
\begin{multline}
\hat{L}_{\mathrm{cm}}=\left(\frac{1}{2\hbar}[(\mathbf{r}_1+\mathbf{r_2})\times(\mathbf{p}_1+\mathbf{p}_2)]\right)_z = {}\\
{}\frac{1}{2}\left(\hat{L}_1+\hat{L}_2+b_1^\dag b_2+b_2^\dag b_1-a_1^\dag a_2-a_2^\dag a_1\right).
\end{multline}

These operators projected onto a single LL have the following form:
\begin{eqnarray}
\hat{L}_{\mathrm{rel}}^p=\frac{1}{2}\left(\hat{L}_1+\hat{L}_2-b_1^\dag b_2-b_2^\dag b_1\right),\\
\hat{L}_{\mathrm{cm}}^p=\frac{1}{2}\left(\hat{L}_1+\hat{L}_2+b_1^\dag b_2+b_2^\dag b_1\right).
\end{eqnarray}

The raising and lowering operators for this "single-level angular momenta"\ are $\hat{b}_1^\dag\mp\hat{b}_2^\dag$ and $\hat{b}_1\mp\hat{b}_2$ respectively. With the help of these operators we can represent the eigenstates of the "single-level angular momenta" in the $(n, S_z)$ Landau level as follows:
\nmq{
|m,M\rangle=\frac{1}{\sqrt{2^{m+M}m!M!}} \times\\ (\hat{b}_1^\dag-\hat{b}_2^\dag)^m (\hat{b}_1^\dag+\hat{b}_2^\dag)^M (\psi_{n,-n})_1(\psi_{n,-n})_2,
}
\begin{eqnarray}
\hat{L}_{\mathrm{rel}}^p|m,M\rangle &=& (m-n)|m,M\rangle,\\
\hat{L}_{\mathrm{cm}}^p|m,M\rangle &=& (M-n)|m,M\rangle.
\end{eqnarray}
Here $M,m \geq 0$. We will say that $|m,M\rangle$ is a state with the relative angular momentum $m$ and the center of mass angular momentum $M$. Since every state has to be antisymmetric under the permutation of the electrons, only states with odd $m$ are present in our Hilbert space. The states $|m,M\rangle$ for $m \in 2\mathbb{Z}_+ + 1$ and $M \in \mathbb{Z}_+$ form a complete orthonormal basis.

Commutation relations of the "angular momenta"\ with the operator $\hat{V} = V(r)$ are
\begin{equation}
\label{comm-rel1st}
\left[\hat{V},\hat{L}_{\mathrm{rel/cm}}^p\right]=0,
\end{equation}
\begin{eqnarray}
\left[\hat{V},\hat{b}_1^\dag+\hat{b}_2^\dag\right]=0 &,& \left[\hat{V},\hat{b}_1+\hat{b}_2\right]=0,\\
\left[\hat{V},\hat{b}_1^\dag-\hat{b}_2^\dag\right] \neq0 &,& \left[\hat{V},\hat{b}_1-\hat{b}_2\right] \neq0.
\label{comm-rellast}
\end{eqnarray}

Thus the interaction potential operator can be represented in the LL as
\begin{equation}
\label{pseudoPotNR}
\hat{V}(r)=\sum_{m,M}|m,M\rangle V_m^{(n,n)}\langle m,M|,
\end{equation}
which solves the two-body problem in the LL.

Matrix elements $V_m^{(n,n)}$ which parametrize the operator are called pseudopotential coefficients (or just pseudopotentials), they were first introduced in \cite{Haldane}. The connection of the pseudopotentials with the interaction potential's matrix elements is obvious since the states $|m,M\rangle$ are orthonormal.

In the following section we shall also need a more general matrix element $V_m^{(n_1,n_2)}$:
\nmq{
\label{Eq_twoElectrState_DiffLLs}
|n_1,n_2,m,M\rangle\rangle=\frac{1}{\sqrt{2^{m+M}m!M!}} \times\\ (\hat{b}_1^\dag-\hat{b}_2^\dag)^m (\hat{b}_1^\dag+\hat{b}_2^\dag)^M (\psi_{n_1,-n_1})_1(\psi_{n_2,-n_2})_2,
}
\nmq{
V_m^{(n_1,n_2)}=\langle\langle n_1,n_2,m,M| \hat{V} |n_1,n_2,m,M\rangle\rangle =\\ \langle\langle n_1,n_2,m,0| \hat{V} |n_1,n_2,m,0\rangle\rangle.
}
It is easy to check that $V_m^{(n_1,n_2)} = V_m^{(n_2,n_1)}$ for the potentials depending on the distance between the electrons only. Note that for technical reasons we do not impose antisymmetry on the wave functions \eqref{Eq_twoElectrState_DiffLLs}, to emphasize this we mark these states with a double bracket $\rangle\rangle$.

For computations, it is often more convenient to express $V_m^{(n_1,n_2)}$ in terms of the potential's Fourier transform \cite{QHE_Book}:
\begin{multline}
\label{potFourier}
\tilde{V}(q) = \frac{1}{l^2} \int d^2r V(r) e^{- i \mathbf{q} \mathbf{r}/l} ={}\\
 {}\frac{2\pi}{l^2} \int_{0}^{\infty}  V(r) J_0(q r/l) r dr ={}\\
{} 2\pi \int_{0}^{\infty}  V(l x) J_0(q x) x dx,
\end{multline}
\begin{equation}
\label{pseudopotFourier}
V_m^{(n_1,n_2)} = \int_{0}^{\infty} \tilde{V}(q) L_{m}(q^2) L_{n_1}(q^2/2) L_{n_2}(q^2/2) e^{-q^2} \frac{q dq}{2 \pi},
\end{equation}
where $J_0$ is the zeroth Bessel function of the first kind, $L_{k}$ are the Laguerre polynomials, $l$ is the magnetic length. Derivation of this formula is presented in Appendix~\ref{FourierFormula}.

Thus, the electrostatic interaction between the electrons located in one Landau level can be expressed through a countable set of pseudopotentials $V_m^{(n,n)}$, where $n$ is the LL number, and $m \in 2\mathbb{Z}_{+} + 1$ is the "relative angular momentum"\ of the two interacting electrons.

\subsection{Many-particle problem}

Here we discuss the problem of many electrons in a Landau level and the numerical diagonalization approach.

Since we know how to express the Hamiltonian of the electron-electron interaction in the Hilbert space of two electrons in a Landau level, we can, in principle, express the many-particle system's Hamiltonian through the pseudopotentials. Typically, such a Hamiltonian cannot be solved analytically, therefore numerical diagonalization is used. A standard complication is that the Hamiltonian is an infinite matrix (since there are an infinite number of orbitals in a Landau level), while numerical diagonalization can only be used for finite matrices.

Therefore, several strategies are used to restrict the system size. One is to restrict the maximum orbital number available to the electrons (i.e. consider only those states in the LL where electrons occupy orbitals with angular momentum quantum number $m \leq m_{\mathrm{max}}$). Such a model is called the "hard cutoff" model for a system on disk. It is also possible to restrict the total angular momentum of the electrons ($\sum_i m_i = M_{\mathrm{max}}$), which gives the "soft cutoff" model for a system on disk \cite{Laughlin_SystemOnDisk, GirvinJach_SystemOnDisk, LaiYu_SystemOnDisk}. The third widely used method is to consider a "system on sphere" \cite{Haldane} (two-dimensional finite sphere with the uniform magnetic field transverse to the sphere is considered instead of plane). A Landau level in a system on sphere is finite (has a finite number of orbitals) from the very beginning, so one does not need to introduce an artificial boundary. The wave functions of the single particle states and pseudopotentials are expressed in a somewhat different way (so the matrix of the Hamiltonian is expressed somewhat differently via spherical pseudopotentials). However, for large enough systems the results on sphere should coincide with the results on disk (since the curvature of the sphere plays little role then). That's why planar pseudopotentials are often used for diagonalization on sphere (see e.g. \cite{DasS}). In this work we do a similar thing: we use diagonalization on sphere with planar pseudopotentials.

The Hamiltonian of a system on sphere/disk is a finite matrix that can be expressed in terms of pseudopotentials introduced in the previous subsection and diagonalized numerically. This enables us to find the spectrum and the eigenstates of the system. A typical thing to do then is to compare the numerically found ground state (and, possibly, the excited states) with some trial wave function to check whether the real state is close to the proposed trial state.\footnote{There is a correspondence between trial states which are proposed for the sphere and for the plane, so a result of the diagonalization on a sphere can be compared with a trial state just in the same way as a result of the diagonlization on a disk.}

There is some peculiarity in choosing the number of electrons and orbitals in the system. If one studies the filling factor $\nu$, then by definition in the thermodynamic limit number of electrons $N_e$ in the LL considered and the number of orbitals available to them $N_{\mathrm{orb}}$ are related by $N_e/N_{\mathrm{orb}} \approx \left\{\nu\right\}$, where $\left\{\nu\right\}$ denotes the fractional part of the filling factor. On the contrary, trial wave functions (as can be seen from examples below) fix the relation between the two numbers not approximately but exactly:
\begin{equation}
\label{Eq_NOrbNElectr}
N_{\mathrm{orb}} = N_e/\left\{\nu\right\} - S + 1.
\end{equation}
Number $S$ is called "shift" and can be different for different trial wave functions.\footnote{The summand $+1$ is for the number of the last available orbital in the zeroth LL $m_{\mathrm{max}}$ in a system on disk to be expressed as $m_{\mathrm{max}} = N_e/\left\{\nu\right\} - S$. This is a commonly used definition of the shift.}. Of course, we expect the properties of the system in the thermodynamic limit to be independent of the precise ratio between the number of electrons and the number of orbitals; but in order to compare an exact state with a trial wave function, the number of orbitals and electrons in each of the two states should be related as described by Eq.~\eqref{Eq_NOrbNElectr}.

So, the procedure of numerical finding the system's ground state and its comparison with trial state is as follows: choose the trial state with which to compare; choose the number of electrons and orbitals in such a way that it corresponds to the trial state; find pseudopotentials; calculate the system's Hamiltonian and diagonalize it; calculate the scalar product of the numerically found ground state with the trial state (the closer it is to 1 the more similar the states are).

Usually the numerical diagonalization can be performed only for relatively small numbers of electrons (around 10 to 20) in most cases. This is far from the thermodynamic limit. However, it has historically tended to be the case that even systems this small can be fairly representative of the thermodynamic limit \cite{QHE_Book, ChakrabortyBook}. Strictly speaking, one should attempt extrapolation to infinite size. However, this is beyond the scope of our project and we do not believe that our main result would be qualitatively changed.

Before proceeding to application of this method to bilayer graphene, we show several examples of trial wave functions in the next sub-subsection.

\subsubsection{Examples of trial wave functions}

Here we consider several examples of trial wave functions in order to understand how they look and how to interpret them (for the purposes of numerical diagonalization). For simplicity, we present the trial wave functions for a system on disk.

The simplest example is a trial wave function for the fully occupied zeroth LL. Let $N$ be the number of electrons which occupy the first $N$ orbitals of the $n = 0$ level. Due to the Pauli principle the only possible state is the Slater determinant of all the occupied single-particle orbitals:
\begin{multline}
\psi(w_1,...,w_N) = \det_{1 \leq i \leq N, 0 \leq m \leq N-1} \psi_{n=0,m}(w_i) \propto{}\\
{} \propto \left|
\begin{array}{cccc}
1       & w_1    & \ldots & w_1^{N-1}\\
1       & w_2    & \ldots & w_2^{N-1}\\
\vdots  & \vdots & \ddots & \vdots\\
1       & w_N    & \ldots & w_N^{N-1}
\end{array}
\right|\times e^{-\sum_i |w_i|^2/4},
\end{multline}
where we used the explicit form of the single-particle wave functions in the zeroth LL \eqref{zerothLLwf}. The determinant on the r.h.s. is the well known Vandermonde determinant. Therefore, we can write down the answer for the wave function, which, up to the normalization constant $\mathcal{N}$, looks as follows:
\nmq{
\psi(w_1,...,w_N) = \mathcal{N} e^{-\sum_i |w_i|^2/4} \prod_{1 \leq i < j \leq N} (w_i - w_j) =\\
\mathcal{N} e^{-\sum_i |w_i|^2/4} P(w_1,...,w_N).
}

This example illustrates the fact that any wave function of electrons in the zeroth LL can be expressed as a polynomial of coordinates $w_i$~--- no $\bar{w}_i$~--- times the Gaussian weight. Below in this sub-subsection instead of the wave function $\psi(w_1,...,w_N)$ we will write out the polynomial part $P(w_1,...,w_N)$. For example, for the fully filled zeroth LL
\begin{equation}
P(w_1,...,w_N) = \prod_{1 \leq i < j \leq N} (w_i - w_j).
\end{equation}

If electrons don't fill the whole LL they will try to keep as big a distance from each other as possible (because of the Coulomb repulsion). Starting from this argument, R.~Laughlin proposed his famous trial wave function for the filling factor $\nu = 1/3$ \cite{Laughlin}:
\begin{equation}
P(w_1,...,w_N) = \prod_{1 \leq i < j \leq N} (w_i - w_j)^3.
\end{equation}
It is easy to convince oneself that it indeed corresponds to $\nu = 1/3$ by counting the number of orbitals used by the electrons in this wave function. It has the shift $S = 3$ (in contrast to the full filling, where $S = 1$). The key idea is that the power 3 significantly reduces the probability of finding two electrons close to each other.

This wave function has been generalized for the fillings $\nu = 1/m$:
\begin{equation}
P(w_1,...,w_N) = \prod_{1 \leq i < j \leq N} (w_i - w_j)^m.
\end{equation}
However, since the wave function of the electrons should be antisymmetric, $m$ has to be odd. So, this wave function can be used only for fillings with odd denominators.

There had not been any need in description of even denominators until the $\nu = 5/2$ FQHE was observed \cite{Willett_EvenDenomQHObserv}. Moore and Read in 1991 proposed their trial wave function for a half-filled LL\footnote{As it has been mentioned already, it is assumed that only the partially filled level is important.} \cite{MooreRead}. The wave function, if written for the zeroth LL, looks like
\begin{multline}
P(w_1,...,w_N) = \text{Pfaff}\left(\frac{1}{w_i-w_j}\right) \prod_{1 \leq i < j \leq N} (w_i - w_j)^2 = {}\\ {}\text{AntiSymm}\left(\frac{1}{w_1-w_2}\frac{1}{w_3-w_4}\ldots\frac{1}{w_{N-1}-w_N}\right)\times{}\\ \prod_{1 \leq i < j \leq N} (w_i - w_j)^2.
\end{multline}
$\text{AntiSymm}\left(\ldots\right)$ denotes the antisymmetrization operator. The antisymmetrized combination which is present here is the Pfaffian of the matrix $M_{ij}$ ($M_{ii} = 0$, $M_{ij} = 1/(w_i-w_j)$). After this expression the Moore-Read wave function is also often called the Pfaffian state. The filling factor associated with this wave function is $\nu = 1/2$, and the shift is $S = 3$. The wave function is evidently antisymmetric.

One can write wave functions for higher Landau levels in a similar explicit fashion, but, in fact, for numerical comparison one only needs the coefficients of the state vector expanded in the basis of orbital occupation numbers. For a zeroth LL wave function those coefficients can be obtained from the polynomial representing it by expanding the polynomial into a linear combination of monomials~--- each $w_i^k$ up to the normalization factor corresponds to an electron occupying the state $\psi_{n=0,m=k}$. One can also interpret the zeroth LL trial wave function as a wave function for a higher LL. For that one should replace $\psi_{0,k} \rightarrow \psi_{n,k-n}$ in the very end of the procedure of getting the coefficients. Therefore, polynomials of electrons' coordinates $w_i$ are used for representing trial wave functions for both zeroth LL and the higher LLs.

Similarly, one can map angular momentum orbitals of the zeroth LL of a planar non-relativistic system to the corresponding angular momentum orbitals of a LL of BLG (or of the system on sphere), so that any antisymmetric polynomial (zeroth LL wave function) can be used as a trial wave function of any LL in BLG or the system on sphere.

Thus, all the trial states, including the Moore-Read Pfaffian, can be used for higher LLs. In the $n = 1$ LL the Moore-Read state's overlap with the numerically found ground state for 12 electrons is close to $0.7$\footnote{By overlap we mean scalar product's absolute value squared.}. This is not as impressive as Laughlin's $98-99\%$, but still remarkably good for the Hilbert space of dimension over 16 thousand (two random vectors would have an overlap near $1/16000$ in such space). The FQHE with $\nu = 5/2$ (which corresponds to a half-filled $n = 1$ LL) is observed in GaAs heterostructures. Numerical studies \cite{Morf_1998_NumericsPfaffian, RezayiHaldane_2000_Pfaffian_Numerics, MorfDasSarma_2002_Pfaffian_Numerics, Feiguin_2008_Pfaffian_DMRG, Simon_2008_BCS_Pfaffian_Numerics} strongly suggest that the state is either the Moore-Read Pfaffian or its particle-hole conjugate called anti-Pfaffian \cite{HalperinRosenow_2007_AntiPfaffian, Nayak_2007_AntiPfaffian}. Recent experimental studies seem to rule out the Pfaffian \cite{Radu_experiment, Ensslin_FQHE_TunnExp}, however, whether the state is anti-Pfaffian remains to be confirmed. One difficulty in dealing with the $\nu = 5/2$ fraction in GaAs is the extreme fragility of the corresponding state.

\subsection{Summary of the section}

In this section we review the basis of numerical diagonalization methodology for non-relativistic FQHE systems: introduce Landau levels, briefly discuss the applicability of the SLLA to the GaAs heterostructures, introduce pseudopotentials, and discuss peculiarities of numerical diagonalization in the non-relativistic systems. We also discuss several examples of trial wave functions, including the Moore-Read Pfaffian, and their representation in the form of holomorphic polynomials of complex coordinates.

\section{Numerical diagonalization approach to the quantum Hall effect in bilayer graphene (single Landau level approximation)}

In this section we discuss peculiarities of the numerical diagonalization method in bilayer graphene in the SLLA. Landau levels in bilayer graphene are introduced, expressions for the pseudopotentials are derived.

\subsection{Bilayer graphene. Hamiltonian of a free electron in bilayer graphene}

Graphene is a one-atom thick layer of graphite, or in other words~--- a two dimensional honeycomb lattice of carbon atoms. Bilayer graphene (BLG) is formed of two layers of graphene (two graphene sheets) with certain matching of lattice points. For a detailed review on graphene and bilayer graphene see Ref.~\cite{Graphene_Review}. We are going to recall only the facts necessary for the following consideration.

The Fermi surface of undoped graphene consists of two points (valleys, usually denoted as $K$ and $K'$) in the first Brillouin zone. One can usually neglect jumping of electrons between the valleys.\footnote{This is due to the fact that jumping needs transfer of a quite big momentum (of the order of $h/a$, where $a$ is the lattice constant and has value around $0.25\ \text{nm}$). For example, matrix element of the Coulomb potential decreases like $1/q$ as the transferred momentum $q$ grows. Therefore, jumping between the valleys is suppressed, with controlling parameter being the ratio of the lattice constant $a$ to the typical spatial scale one is interested in. (In our case this is the magnetic length $l$; typical values of the magnetic length are around $l = 10\ \text{nm}$).} Therefore, the valley index of an electron is a good quantum number.

In BLG the low-energy excitations are also located near the same Fermi points in the momentum space. The low-energy BLG Hamiltonian (without external magnetic field) can then be written as \cite{Graphene_Review, MCF}\footnote{The Hamiltonian is written in the basis corresponding to the atomic sites $A$, $\tilde{B}$, $\tilde{A}$, $B$ in the K valley and $\tilde{B}$, $A$, $B$, $\tilde{A}$ in the $K'$ valley. The sites $A$ and $B$ are situated in the bottom graphene layer, while $\tilde{A}$ and $\tilde{B}$ are in the top layer. Our convention is the same as the one used in Refs.~\cite{Graphene_Review, MCF} except for a redefinition of $U$.}
\begin{equation}
\label{Ham4}
H_{1-\mathrm{part}}^{\mathrm{BLG}} = \xi \left(
\begin{array}{cccc}
-U & 0 & 0 & v \pi^\dag\\
0 & U& v\pi& 0\\
0 & v\pi^\dag& U &\xi \gamma_1\\
v\pi & 0 &\xi\gamma_1 & -U
\end{array}
\right),
\end{equation}
where $\xi=\pm1$ is the valley index such that $\xi = +1$ corresponds to $K$ and $\xi = -1$ corresponds to $K'$, $\pi = p_x+ i p_y$ is the complex momentum. The spectrum has a mini-gap $2 U$, which can be tuned by the external electric field perpendicular to the bilayer graphene sheet.\footnote{One can think that the electrostatic potential of one layer is $U$, while the other layer's potential is $-U$.} We will call $U$ the "mini-gap parameter". The Fermi velocity is taken to be $v \approx 10^6\ \text{m/s}$, and the interlayer hopping constant is taken to be $\gamma_1 \approx 0.35\ \text{eV}$ \cite{Graphene_Review}.

In the absence of the external electric field (when $U = 0$) the low-energy spectrum has quadratic form $E = \pm |\pi|^2/(2 m^*)$, where the effective mass $m^* = \gamma_1/(2 v^2) \approx 0.03 m_e$ \cite{MCF}.

\subsection{Problem of a single BLG electron in a magnetic field}

The Hamiltonian of an electron in bilayer graphene in the perpendicular magnetic field is obtained by making the derivatives covariant and taking the spin energy into account:
\begin{equation}
\label{Ham4}
H_{1-\mathrm{part}}^{\mathrm{BLG}} = \xi \left(
\begin{array}{cccc}
-U & 0 & 0 & v \pi^\dag\\
0 & U& v\pi& 0\\
0 & v\pi^\dag& U &\xi \gamma_1\\
v\pi & 0 &\xi\gamma_1 & -U
\end{array}
\right)
- g \mu_{\mathrm{B}} B S_{z},
\end{equation}
where $\xi=\pm1$ is for the two valleys, $\pi = \pi_x+ i \pi_y$ (see definition of $\pi_i$ after formula \eqref{H_NR}). In BLG the Lande factor $g \approx 2$. Without loss of generality we will consider only the case $B, U > 0$.

It is easy to express the complex momenta through the operators (\ref{ab-operatorsb}-\ref{ab-operatorse}):
\begin{equation}
\pi = - i \sqrt{2} \hbar l^{-1} \hat{a},\quad \pi^{\dag} = i \sqrt{2} \hbar l^{-1} \hat{a}^{\dag}.
\end{equation}

Thus the Hamiltonian can be expressed as
\begin{equation}
\label{Ham4transf}
H_{1-\mathrm{part}}^{\mathrm{BLG}} = \xi \hbar \omega_c \left(
\begin{array}{cccc}
-u & 0 & 0 & i \gamma \hat{a}^\dag\\
0 & u& -i \gamma \hat{a}& 0\\
0 & i \gamma \hat{a}^\dag& u &\xi \gamma^2\\
-i \gamma \hat{a} & 0 &\xi\gamma^2 & -u
\end{array}
\right)
- g \mu_{\mathrm{B}} B S_{z},
\end{equation}
where we introduced $\omega_c = e B/(m^* c) = 2 v^2 e B/\gamma_1 c$ (after definition of ref.~\cite{MCF}), $\gamma^2 = \gamma_1/(\hbar \omega_c)$, $u = U/(\hbar \omega_c)$.

This Hamiltonian does not commute with the $z$-projection of the orbital angular momentum $\hat{L}$ defined in Eq.~\eqref{Ldef}. We introduce the $z$-projection of the "pseudospin angular momentum"\ $\hat{\Sigma}$:
\begin{equation}
\hat{\Sigma} = \left(
\begin{array}{cccc}
1 & 0 & 0 & 0\\
0 & -1& 0 & 0\\
0 & 0 & 0 & 0\\
0 & 0 & 0 & 0
\end{array}
\right).
\end{equation}
Then the $z$-projection of the full angular momentum $\hat{J} = \hat{L} + \hat{\Sigma}$ does commute with the Hamiltonian.

Now it is easy to express the general form of the spatial part of the Hamiltonian's \eqref{Ham4transf} eigenstates through the non-relativistic wave functions (\ref{NRwfb}-\ref{NRwfe}):
\begin{equation}
\Psi_{n m} = \left(
\begin{array}{c}
A_{n} \psi_{n m}\\
B_{n} \psi_{n-2, m+2}\\
C_{n} \psi_{n-1, m+1}\\
D_{n} \psi_{n-1, m+1}
\end{array}\right),
\end{equation}
the sense of the number $n$ is similar to the Landau level number, while $m$ corresponds to the projection of the full angular momentum $j_z = m + 1$. The amplitudes $A_{n}, B_{n}, C_{n}, D_{n}$ do not depend on $m$.

Acting on this wave function with the Hamiltonian and demanding it to be an eigenfunction we find the equation for the eigenvalues:
\begin{equation}
\label{spectreq}((u-\xi \varepsilon)^2-\gamma^2 n)((u+\xi \varepsilon)^2-\gamma^2 (n-1)) = \gamma^4 (\varepsilon^2-u^2)
\end{equation}
where $\varepsilon = (E + 2 \mu_B B S_z)/(\hbar \omega_c)$, and $E$ is the energy.

Finding the single particle spectrum for the realistic values of the parameters, we see that the levels split into two groups: the one with $|E| < \gamma_1$ and the one with $|E| \geq \gamma_1$. The Zeeman splitting is negligibly small, just like the non-relativistic case. The levels with $|E| < \gamma_1$, which we are interested in can be characterized by five quantum numbers: the valley index $\xi$, the LL number $n \in \mathbb{Z}_{+}$, the full angular momentum projection $j_z = m + 1$ with $m \in (\mathbb{Z}_{+}-n)$, $s = \pm1$ (which shows whether the energy is positive or negative) and $S_z = \pm 1/2$. Thus the wave functions (their spatial components) in the $n$-th LL look like
\begin{equation}
\label{wavefunction}
\Psi_{n m}^{\xi s} = \left(
\begin{array}{c}
A_n^{\xi s} \psi_{n m}\\
B_{n}^{\xi s} \psi_{n-2, m+2}\\
C_{n}^{\xi s} \psi_{n-1, m+1}\\
D_{n}^{\xi s} \psi_{n-1, m+1}
\end{array}\right).
\end{equation}
The amplitudes which are present in this formula can be expressed as follows
\begin{gather}
A_n^{\xi s} = \mathcal{N},\\
B_{n}^{\xi s} = - \frac{\sqrt{n-1}}{u+\xi \varepsilon_{n}^{\xi s}} \frac{(u-\xi \varepsilon_{n}^{\xi s})^2-\gamma^2 n}{\xi \gamma^2 \sqrt{n}} \mathcal{N},\\
C_{n}^{\xi s} = -i \frac{(u-\xi \varepsilon_{n}^{\xi s})^2-\gamma^2 n}{\xi \gamma^3 \sqrt{n}} \mathcal{N},\\
D_{n}^{\xi s} = i \frac{u-\xi \varepsilon_{n}^{\xi s}}{\gamma \sqrt{n}} \mathcal{N},
\end{gather}
where $\mathcal{N}$ is a normalization constant. Obviously, they depend on the magnetic field $B$ and the mini-gap parameter $U$.

\begin{figure}[tbp]
  \centering
  \includegraphics[width=.96\columnwidth]{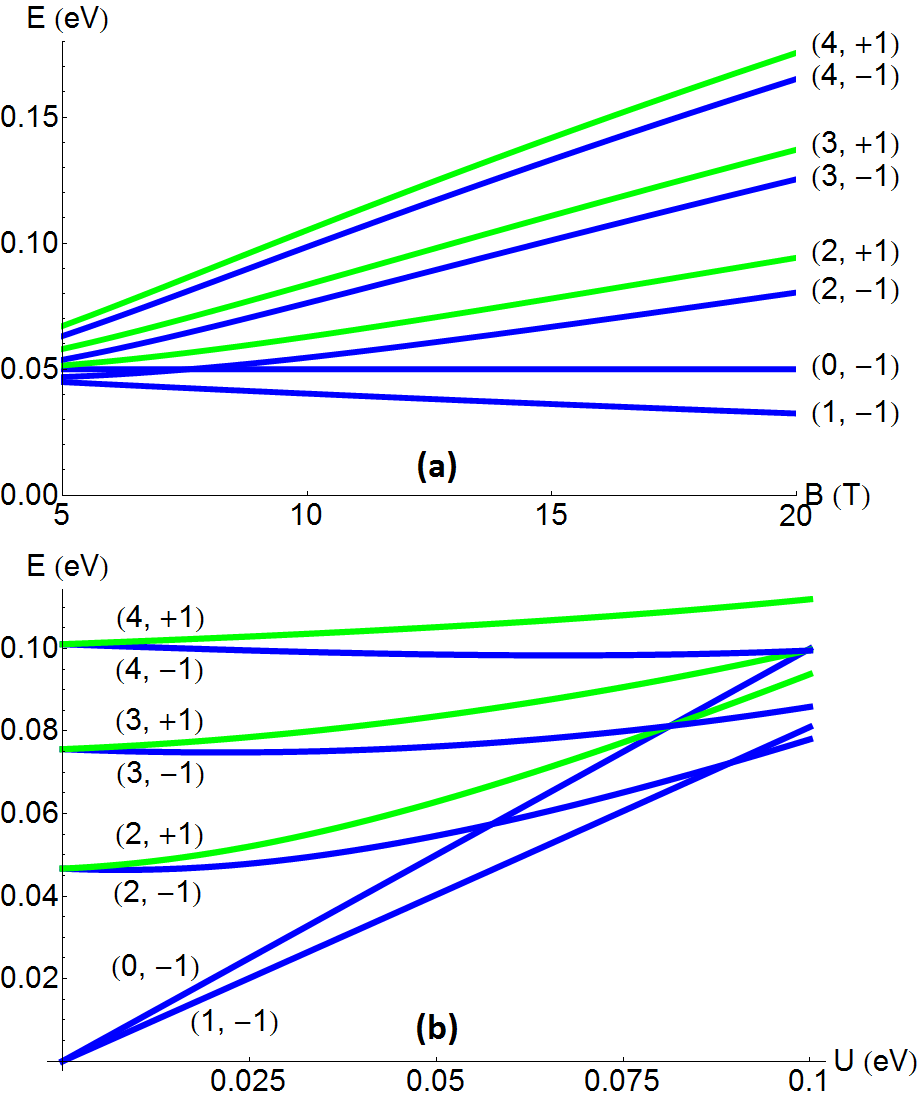}
  \caption{\textbf{Dependence of the lowest LLs' energies on (a) the magnetic field $B$ for $U = 50\ \text{meV}$, and (b) on the mini-gap parameter $U$ for $B = 10\ \text{T}$.} Each level is labeled by a pair of quantum numbers $(n, \xi)$. Only positive-energy part of the spectrum is shown. (Color online).}
  \label{fig:spectr}
\end{figure}

Before considering the two particle problem in bilayer graphene, we have a look at the single-particle spectrum.
Fig.~\ref{fig:spectr}a shows the dependence of the several lowest Landau levels on the magnetic field for $U = 50\ \text{meV}$. Only the positive part of the spectrum is shown, the negative part can be obtained with the help of electron-hole conjugation ($\varepsilon_{n}^{\xi, -s} = -\varepsilon_{n}^{-\xi, s}$). Each positive LL is labeled by a pair of quantum numbers $(n, \xi)$. When we need to work with negative LLs, we label them with $j = (n, \xi, s)$. One can see that for large values of the magnetic field the levels form quasidegenerate doublets which are separated by energies of the order of $\hbar \omega_c$. Fig.~\ref{fig:spectr}b shows the dependence of the same LLs' energies on the mini-gap parameter $U$ for the magnetic field $B = 10\ \text{T}$. Note that for large enough values of $U$ (or for small enough values of $B$) multiple crossings of Landau levels occur. It is easy to understand that when several LLs are close to each other (for small magnetic fields/large mini-gaps when the LLs cross, or for small values of the mini-gap when the levels in a doublet are almost degenerate) significant deviation from the SLLA can occur. Thus, the applicability of the SLLA puts constraints onto the external parameters. This point is discussed in details in section \ref{Sect_DeviationsFromSLLA}.

\subsection{\label{2pBLG}Interaction of two electrons in a Landau level of bilayer graphene}
The two-particle problem within the SLLA in BLG can be solved analogously to the non-relativistic case.

Define the projections of the relative and the center of mass full angular momenta to a single Landau level:
\nmq{
\label{BLG_Jproj}
\hat{J}_{\mathrm{rel}/\mathrm{cm}}^p=\hat{L}_{\mathrm{rel}/\mathrm{cm}}^p+\frac{1}{2}(\Sigma_1+\Sigma_2) =\\ \frac{1}{2}(\hat{J}_1+\hat{J}_2\mp (b_1^\dag b_2+b_2^\dag b_1)).
}

The commutation relations of the projected angular momenta and their raising and lowering operators with the parts of the two-particle Hamiltonian are similar to the non-relativistic case:
\begin{eqnarray}
\hat{H}_{2-\mathrm{part}}^{\mathrm{BLG}} &=& \hat{H}_{\mathrm{free}}^{\mathrm{BLG}}+V(r),\\
\hat{H}_{\mathrm{free}}^{\mathrm{BLG}} &=& \hat{H}_{1-\mathrm{part},1}^{\mathrm{BLG}}+\hat{H}_{1-\mathrm{part},2}^{\mathrm{BLG}},
\end{eqnarray}
\begin{eqnarray}
\left[\hat{H}_{\mathrm{free}}^{\mathrm{BLG}},\hat{J}_{\mathrm{rel}/\mathrm{cm}}^p\right]=0 &,& \left[\hat{H}_{\mathrm{free}}^{\mathrm{BLG}},\hat{b}_1^\dag\mp\hat{b}_2^\dag\right]=0,\\ \left[\hat{H}_{\mathrm{free}}^{\mathrm{BLG}},\hat{b}_1\mp\hat{b}_2\right]=0 &,& \left[\hat{V},\hat{J}_{\mathrm{rel}/\mathrm{cm}}^p\right]=0,\\ \left[\hat{V},\hat{b}_1^\dag+\hat{b}_2^\dag\right]=0 &,& \left[\hat{V},\hat{b}_1+\hat{b}_2\right]=0,\\
\left[\hat{V},\hat{b}_1^\dag-\hat{b}_2^\dag\right] \neq0 &,& \left[\hat{V},\hat{b}_1-\hat{b}_2\right] \neq0.
\end{eqnarray}

The simultaneous eigenstates of the two "angular momenta" \eqref{BLG_Jproj} in a Landau level have the form
\nmq{
|m,M\rangle=\frac{1}{\sqrt{2^{m+M}m!M!}} \times\\ (\hat{b}_1^\dag-\hat{b}_2^\dag)^m(\hat{b}_1^\dag+\hat{b}_2^\dag)^M(\Psi_{n,-n}^{\xi s})_1(\Psi_{n,-n}^{\xi s})_2,
}
\begin{eqnarray}
\hat{J}_{rel}^p|m,M\rangle &=& (m-n+1)|m,M\rangle,\\
\hat{J}_{cm}^p|m,M\rangle &=& (M-n+1)|m,M\rangle.
\end{eqnarray}

Thus, just like in a non-relativistic system, the two-particle interaction potential in a Landau level can be expressed through pseudopotentials:
\begin{equation}
\hat{V}(r)=\sum_{m,M}|m,M\rangle V_m^{n, \xi, s}\langle m,M|.
\end{equation}
The expression for the pseudopotentials in terms of matrix elements of the potential is straightforward as the states $|m,M\rangle$ are orthonormal. These pseudopotentials can be expressed via the non-relativistic pseudopotentials:
\begin{multline}
\label{BLGppviaNRpp}
V_m^{n, \xi, s} = |A_{n}^{\xi s}|^4 V_m^{(n,n)} + |B_{n}^{\xi s}|^4 V_m^{(n-2,n-2)} +{}\\
 {}(|C_{n}^{\xi s}|^2 + |D_{n}^{\xi s}|^2)^2 V_m^{(n-1,n-1)} +{}\\
  {}2 |A_{n}^{\xi s}|^2 |B_{n}^{\xi s}|^2 V_m^{(n,n-2)} +{}\\
   {}2 |A_{n}^{\xi s}|^2 (|C_{n}^{\xi s}|^2 + |D_{n}^{\xi s}|^2) V_m^{(n,n-1)} +{}\\
    {} 2 |B_{n}^{\xi s}|^2 (|C_{n}^{\xi s}|^2 + |D_{n}^{\xi s}|^2) V_m^{(n-2,n-1)}.
\end{multline}
This expression makes obvious the possibility of tuning of the pseudopotentials by changing the values of the amplitudes $A_{n}^{\xi s}, B_{n}^{\xi s}, C_{n}^{\xi s}, D_{n}^{\xi s}$. Since the amplitudes depend on the external parameters $U$ and $B$, the pseudopotentials can be tuned with the help of external perpendicular electric and magnetic fields.

For practical purposes it is useful to incorporate formula \eqref{pseudopotFourier} into Eq.~\eqref{BLGppviaNRpp} which leads to
\begin{equation}
\label{ppFourierBLG}
V_m^{n, \xi, s} = \int_{0}^{\infty} \tilde{V}(q) L_{m}(q^2) \left(F_{n}^{\xi s}(q^2/2)\right)^2 e^{-q^2} \frac{q dq}{2 \pi},
\end{equation}
\begin{multline}
\label{FormFactorBLG}
F_{n}^{\xi s}(q^2/2) = |A_{n}^{\xi s}|^2 L_{n}(q^2/2) + |B_{n}^{\xi s}|^2 L_{n-2}(q^2/2) +{}\\
 {}(|C_{n}^{\xi s}|^2 + |D_{n}^{\xi s}|^2) L_{n-1}(q^2/2).
\end{multline}

After pseudopotentials are calculated the numerical diagonalization procedure is identical to the non-relativistic case\footnote{Recall that though the trial wave functions are written in the form of complex polynomials, they, in fact, give decomposition of the wave function into Slater determinants of the single particle states. Thus they are applicable to any system with Landau levels having structure similar to the non-relativistic case, so they are applicable to the BLG. If a trial state is written in the basis of occupation numbers the only difference to the diagonalization and comparison procedure is that one has to use the BLG pseudopotentials instead of the non-relativistic ones.}.

We emphasize that the analysis in this section is done strictly within the SLLA. As is discussed in the next section, in BLG the SLLA is less justified than in GaAs systems. There are, however, conditions under which the SLLA is a good approximation. In the latter case the effects of other LLs can be incorporated into the SLLA by means of perturbation theory.

\subsection{Summary of the section}

In this section the explicit formulae for the SLLA in BLG are provided (wave functions in a Landau level, expression for the pseudopotentials). It is shown that application of the numerical diagonalization methodology to BLG system within the SLLA does not differ too much from the application to a non-relativistic system.

\section{\label{Sect_DeviationsFromSLLA}Deviations from the single Landau level approximation in bilayer graphene}

As discussed in the previous section, studying the FQHE in BLG within the SLLA does not differ too much from the non-relativistic case. However, as it has already been mentioned the single-particle spectrum of BLG restricts the applicability of the SLLA, and under the usual experimental conditions the restriction is significant. Depending on several factors, there are three regimes:
\begin{enumerate}
  \item the SLLA is fully reliable;
  \item the SLLA is completely inapplicable because one has to consider several Landau levels together since the electrons partially occupy each of those;
  \item intermediate regime, the effects of the presence of other Landau levels can be incorporated into the SLLA as corrections to the intra-level electron-electron interaction.
\end{enumerate}

Note that the perturbatively small corrections to the electron-electron interaction in the intermediate regime may lead to a qualitative change in the phase diagram. Indeed, it is known that small corrections to the pseudopotentials' values can lead to a transition to a different topological phase or to a collapse of the bulk gap (see, e.g., Ref.~\cite{DasS}).

In this section we discuss factors which determine boundaries between the regimes, we also show how to take into account the corrections in the third regime. First, a brief discussion is presented in subsection~ \ref{ImpBLGeffBrief}, then the technical details are given in subsection~\ref{ImpBLGeffDet}.

\subsection{\label{ImpBLGeffBrief}Effects important in bilayer graphene}

Now we discuss in details the effects which are important in BLG. For the reasons discussed earlier, we continue suppressing spin quantum numbers of electrons.

The important effects are as follows:
\begin{itemize}
  \item Firstly, BLG in perpendicular electric field is a narrow gap semiconductor, thus the effects of vacuum polarization are strong \cite{Aleiner2012}.
  \item Secondly, Coulomb interaction of electrons can lead to mixing of Landau levels, or to emergence of spin or/and valley unpolarized states. Coulomb interaction can also lead to intervalley hopping of electrons. Even though such processes are suppressed compared to intravalley scattering one should still estimate their relevance.
  \item Thirdly, even when LL mixing is small, virtual hopping between the LLs can change (renormalize) intra-level electron-electron interaction.
\end{itemize}
Next we discuss each of those effects in detail.

\subsubsection{Vacuum polarization}

\begin{figure}[tbp]
  \centering
  \includegraphics[width=.96\columnwidth]{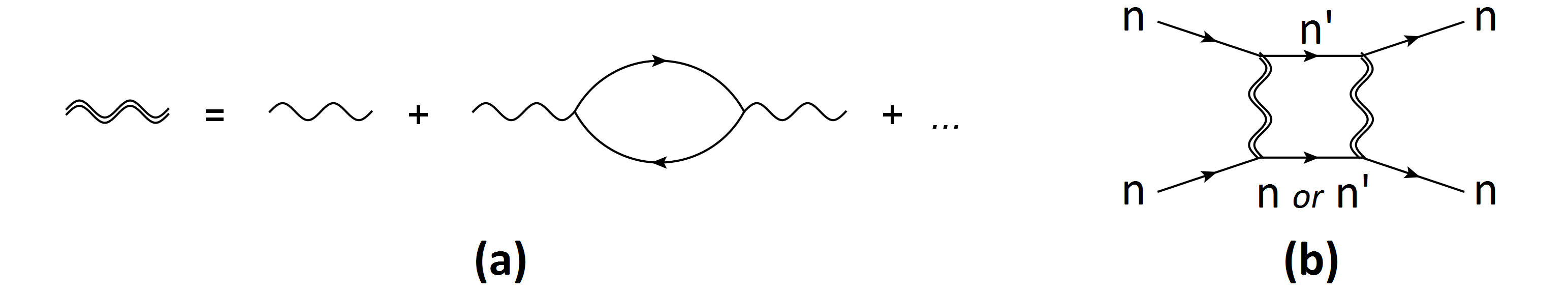}
  \caption{\textbf{Feynman diagrams showing renormalization of the electron-electron interaction} due to (a) the vacuum polarization processes, and (b) the simplest processes involving virtual hopping of one or both of the two interacting electrons from the $n$-th LL to the $n'$-th LL.}
  \label{fig:diag}
\end{figure}

The virtual processes shown in Fig.~\ref{fig:diag}a lead to renormalization of the electron-electron interaction. This is important since the interaction determines the FQHE. The Fourier transform of the renormalized (screened) interaction potential can be expressed as\footnote{A similar screening approach has been discussed in GaAs;
see, for example, \cite{Morf_Screening}.}
\begin{equation}
\label{screening}\tilde{V}_{\mathrm{scr}}(q) = \frac{\tilde{V}(q)}{1 + l^2 \tilde{V}(q) \Pi(q, \omega = 0)}
\end{equation}
where $\tilde{V}(q) = 2\pi e^2/(l q \kappa)$ is the Fourier transform of the unscreened Coulomb potential, $\kappa$ is the dielectric constant, which is felt by the system's electrons\footnote{The dielectric constant sensed by BLG is $\kappa = (\kappa_1+\kappa_2)/2$, where $\kappa_1$ and $\kappa_2$ are the dielectric constants of the environment below and above the sheet of BLG.}, and $\Pi(q, \omega)$ is the polarization function. Since we are interested in the effects at the energy scales much less than the inter-LL distances we can neglect the retardation effects (use only $\omega = 0$).

We compute the polarization function for the BLG in magnetic field within the RPA (random phase approximation), which can be justified within the $1/N$-expansion \cite{Aleiner2012} ($N = 2 \text{ spin projections} \times 2 \text{ valleys} = 4$). Since $\Pi(q, \omega = 0) \propto q^2$, screening is not efficient at large distances; however, it strongly affects the first few Haldane pseudopotentials (corresponding to distances of the order of the magnetic length) which have the most significant impact on the stability of any FQHE state. Details of the calculation are described in sub-subsection \ref{PolFuncDet}.

\subsubsection{\label{PopRevGen}Landau level mixing and population reversal}

The order of levels in Fig.~\ref{fig:spectr} prescribes the natural order of filling of the LLs by electrons in the independent electron approximation. However, it can happen that for some filling fractions the electron-electron interaction leads to a reversal of this natural order in a part of parameter space (by external parameters we mean the magnetic field, the mini-gap and the dielectric constant). For example, the Coulomb energy of the fully filled $(2, +1)$ LL is less than the one of the fully filled $(2, -1)$ for $U > 0$. In the region where the interaction is strong compared to the gap between the two levels this can lead to the fully filled $(2, +1)$ LL having lower total energy than the fully filled $(2, -1)$ level. Thus, the former would be filled before the latter.

Whereas for fractional filling such effects are much more difficult to analyze, population reversal at the integer filling fraction would be an indicator of a strong violation of the SLLA. Thus, we constrain our analysis to the region of the parameter space where no population reversal occurs at integer filling. More details on the population reversal issues are presented in sub-subsection \ref{PopRevDet}.

When the quasidegenerate levels are from different valleys, valley-unpolarized states may be preferred, particularly for fractional filling. Furthermore, when the quasidegenerate levels are from the same valley (as in the $n = 0$ and $n = 1$ case) level mixing may occur. These are interesting effects which are, however, beyond the scope of this work.

In any of the cases (valley unpolarized state or level mixing) one has to consider several LLs simultaneously. This is hard technically since for the same number of electrons the system's Hilbert space is significantly larger, which complicates use of the numerical diagonalization. Therefore, we restrict study to the region of the parameter space where no valley unpolarized states and no level mixing occur. Our criteria for smallness of level mixing and valley unpolarization are discussed in sub-subsection \ref{PopRevDet}.

For the valley-polarized states one can still investigate whether the state is spin polarized. Generally, spin-unpolarized states are not favored by Coulomb repulsion unless the pseudopotential is hollow core (pseudopotential do not fall off monotonically with relative angular momentum). We find that the screened pseudopotential does fall off monotonically for $\kappa \gtrsim 10$ in all the cases considered in section~\ref{PossObsPfaffBLG}. For $\kappa \lesssim 10$ the pseudopotential is weakly non-monotonic in some regions of the parameter space. We, therefore, consider only spin-polarized states without further investigation of the restrictions that this condition imposes on the parameter space.

\subsubsection{Renormalization of pseudopotentials due to virtual hopping}

The SLLA is exact in the limit of infinite energy difference $\Delta E$ between the LLs. For finite $\Delta E$ the pseudopotentials acquire corrections due to virtual transitions between the LLs such as, for example, shown in Fig.~\ref{fig:diag}b. Such corrections are theoretically tractable only in the perturbative regime (when they are small); however, even the presence of small corrections may dramatically affect the phase diagram due to the extreme sensitivity of the FQHE states to the details of the interaction (see e.g. \cite{DasS}).

We restrict the region of validity of our consideration by requiring the typical interaction energy scale (it can be interaction potential value at the magnetic length distance or, almost equivalently, the zeroth pseudopotential) to be smaller than the distances to each of the neighbouring LLs from the same valley. In these regions we take into account corrections to the pseudopotentials up to second order perturbation theory (Fig.~\ref{fig:diag}b).

The second order perturbation theory also gives rise to three-particle interaction \cite{SimonRezayi_LLmixing_perturb, PetersonNayak_LLmixing_perturb, SodemannMacDonald_LLmixing_perturb, PetersonNayak2_LLmixing_perturb}, which can play an important role in some cases. For example, the three-body interaction is crucial to distinguish between the Moore-Read Pfaffian and anti-Pfaffian. However, in the present work we neglect these terms.

The details of the calculation of the corrections are presented in sub-subsection~\ref{hoppingQM}.

\subsubsection{General plan for numerical study of a fractional QHE in bilayer graphene}

With the remarks made above, the general plan for study of FQHE at a certain filling fraction on a certain LL can be formulated as follows:
\begin{enumerate}
  \item Calculate the screened interaction potential.
  \item Determine the region of parameter space in which no valley unpolarized states emerge and level mixing doesn't take place\footnote{By external parameters we mean the magnetic field, the mini-gap parameter and the dielectric constant.}.
  \item Calculate pseudopotentials in this region of parameter space. Take the corrections due to virtual hopping into account (the modified SLLA).
  \item Use the calculated corrected pseudopotentials for numerical diagonalization and compare the exact numerically found ground state with the trial one.
\end{enumerate}

The next subsection contains details of calculation of the polarization function, of the corrections to the pseudopotentials, of the criterion for absence of population reversal of LLs, and of the criteria for absence of valley unpolarization and LL mixing.

\subsection{\label{ImpBLGeffDet}Effects important in bilayer graphene: calculation details}

\subsubsection{\label{PolFuncDet}Calculation of the polarization function}

Here we calculate the vacuum polarization function. We first do this for the case of integer filling. At the end of the sub-subsection we generalize this calculation to the case of a fractional filling factor.

The polarization function within the RPA is just a density-density correlation function\footnote{Density is meant to be normal ordered: electron/hole creation operators should be to the left of the annihilation operators.} in the free theory (this corresponds to the fermionic loop in Fig.~\ref{fig:diag}a)
\begin{equation}
\Pi(\mathbf{r}-\mathbf{r}',t-t') = - i \langle T \rho(\mathbf{r},t) \rho(\mathbf{r}',t') \rangle,
\end{equation}
$\hbar$ is put to be $1$ in this sub-subsection, the $T$-symbol denotes time ordering:
\begin{equation}
T \rho(\mathbf{r},t) \rho(\mathbf{r}',t') =
\left\{
\begin{array}{c}
\rho(\mathbf{r},t) \rho(\mathbf{r}',t'), t>t'\\
\rho(\mathbf{r}',t')  \rho(\mathbf{r},t), t<t'
\end{array}
\right. .
\end{equation}

The density-density correlator is translation-invariant since the system is uniform, so
\begin{equation}
\langle T \rho(\mathbf{r},t) \rho(\mathbf{r}',t') \rangle = \langle T \rho(\mathbf{r}-\mathbf{r}',t-t') \rho(\mathbf{0},0) \rangle.
\end{equation}

Let's denote the set of quantum numbers $(n, m, \xi, s)$ by $k$, and write $k < k_F$ if the state is occupied, and $k > k_F$ otherwise. The polarization function can be expressed with the help of the wave functions \eqref{wavefunction} in the following way:
\begin{widetext}

\begin{multline}
\Pi(\mathbf{r},t) = - 2i\left[\sum_{k < k_F, k' > k_F, \xi=\xi'}\Psi_k(x)^{\dag} \Psi_{k'}(x) \Psi_{k'}(0)^{\dag} \Psi_{k}(0) \times e^{i (E_k-E_{k'}) t} \theta(t) +\right.{}\\
 {}\left.\sum_{k > k_F, k' < k_F, \xi=\xi'}\Psi_k(x)^{\dag} \Psi_{k'}(x) \Psi_{k'}(0)^{\dag} \Psi_{k}(0)\times e^{i (E_k-E_{k'}) t} \theta(-t)\right],
\end{multline}
\begin{equation}
\theta(x) = \left\{
\begin{array}{c}
1, x>0\\
0, x<0
\end{array}
\right. .
\end{equation}
The factor of $2$ in front of the square brackets is due to spin.

The Fourier transform of the polarization function is then defined as
\begin{equation}
\label{PolFunc_Fourier}
\Pi(\mathbf{q}, \omega) = \int d^2r dt e^{-i (\mathbf{q} \mathbf{r}/l - \omega t)} \Pi(\mathbf{r}, t).
\end{equation}
So the polarization function at the zero frequency $\Pi(\mathbf{q}, \omega = 0)$, which we need to find the interaction potential, can be expressed as
\eq{
\Pi(\mathbf{q},\omega = 0) = 2 \sum_{k > k_F, k' < k_F, \xi=\xi'} \frac{1}{E_k - E_{k'}} \int d^2r e^{-i \mathbf{q} \mathbf{r}/l} \left(\Psi_k(x)^{\dag} \Psi_{k'}(x) \Psi_{k'}(0)^{\dag} \Psi_{k}(0) + c.c.\right).
}

After a short calculation we find that
\eq{
\Pi(\mathbf{q},\omega = 0) =
 \frac{2}{l^2} \sum_{j > j_F, j' < j_F, \xi=\xi'} \frac{1}{E_j - E_{j'}} \times (I_{n,s,n',s'}+c.c.) =
  \frac{1}{l^2 \hbar \omega_c} \Pi_{\mathrm{dimless}}(q),
}
\begin{multline}
I_{n,s,n',s'} = (-1)^{(n-n')} \int_{0}^{\infty} dr\ r J_{0}(q r / l) \times \left[ A_n^{\xi s} \overline{A_{n'}^{\xi s'}} \left(\overline{A_n^{\xi s}} A_{n'}^{\xi s'} \overline{\psi_{n,0}(w)}\psi_{n',0}(w) +\right. \right. {}\\
{}\left. \overline{B_n^{\xi s}} B_{n'}^{\xi s'} \overline{\psi_{n-2,2}(w)}\psi_{n'-2,2}(w)+ (\overline{C_n^{\xi s}} C_{n'}^{\xi s'}+\overline{D_n^{\xi s}} D_{n'}^{\xi s'}) \overline{\psi_{n-1,1}(w)}\psi_{n'-1,1}(w)\right) + {}\\
 {}B_n^{\xi s} \overline{B_{n'}^{\xi s'}} \left(\overline{A_n^{\xi s}} A_{n'}^{\xi s'} \overline{\psi_{n,-2}(w)}\psi_{n',-2}(w) + \overline{B_n^{\xi s}} B_{n'}^{\xi s'} \overline{\psi_{n-2,0}(w)}\psi_{n'-2,0}(w) +\right.{}\\
  {}\left. (\overline{C_n^{\xi s}} C_{n'}^{\xi s'}+\overline{D_n^{\xi s}} D_{n'}^{\xi s'}) \overline{\psi_{n-1,-1}(w)}\psi_{n'-1,-1}(w)\right) + {}\\
 {}(C_n^{\xi s} \overline{C_{n'}^{\xi s'}}+D_n^{\xi s} \overline{D_{n'}^{\xi s'}}) \left(\overline{A_n^{\xi s}} A_{n'}^{\xi s'} \overline{\psi_{n,-1}(w)}\psi_{n',-1}(w) + \overline{B_n^{\xi s}} B_{n'}^{\xi s'} \overline{\psi_{n-2,1}(w)}\psi_{n'-2,1}(w)\right.{}\\
  {}\left.\left. + (\overline{C_n^{\xi s}} C_{n'}^{\xi s'}+\overline{D_n^{\xi s}} D_{n'}^{\xi s'}) \overline{\psi_{n-1,0}(w)}\psi_{n'-1,0}(w)\right)\right],
\end{multline}
where $j$ denotes a set $(n, \xi, s)$, overbar denotes complex conjugation.\footnote{Note that for some values of $n$, $n'$ this general expression has the non-defined wave functions like, e.g., $\psi_{n'-2,0}(w)$ for $n' = 0$ or $n' = 1$. These terms, however, do not contribute, which is ensured by the coefficients $B_{n'}^{\xi s'}$, $C_n^{\xi s}$ etc. which take zero values in those cases.} The integrals in the expression above can be found analytically:
\eq{
\int_{0}^{\infty} dr\ r J_{0}(q r / l) \overline{\psi_{n,m}(w)}\psi_{n',m}(w) =
 \frac{1}{2\pi} \int d^{2}r \overline{\psi_{n,m}(w)} e^{i \mathbf{q} \mathbf{r}/l} \psi_{n',m}(w) =
  \frac{1}{2\pi} F_{n, n'}(x) F_{n+m, n'+m}(\bar{x}) e^{-q^2/2},
}
\eq{
F_{i_1,i_2}(x)|_{i_1 \geq i_2} =
 \sum_{k = 0}^{i_2} \frac{\sqrt{i_1! i_2!}}{k! (k+i_1-i_2)! (i_2 - k)!} \left(-\frac{|x|^2}{2}\right)^k \left(\frac{i x}{\sqrt{2}}\right)^{i_1-i_2} =
  \sqrt{\frac{i_2!}{i_1!}} \left(\frac{i x}{\sqrt{2}}\right)^{i_1-i_2} L_{i_2}^{(i_1-i_2)}(|x^2|/2),
}
\begin{equation}
F_{i_1,i_2}(x)|_{i_1 \leq i_2} = F_{i_2,i_1}(\bar{x}),
\end{equation}
where $x = q_x + i q_y$, and $L_{n}^{(\alpha)}$ are generalized Laguerre polynomials. The derivation is very similar to the derivation of formula \eqref{pseudopotFourier} presented in Appendix~\ref{FourierFormula}.
\end{widetext}

The Fourier transform of the interaction potential \eqref{potFourier}, \eqref{screening} can be expressed then as
\begin{equation}
\label{screening_dimless}\tilde{V}_{\mathrm{scr}}(q) = \frac{2 \pi e^2}{\kappa q l}\frac{1}{1 + 2 \pi \frac{e^2}{l \kappa \hbar \omega_c} \frac{\Pi_{\mathrm{dimless}}(q)}{q}}.
\end{equation}

In the case of a fractional filling factor we have to take into account not only fully filled or entirely empty LLs but the partially filled ones as well. We do this with the help of the following approximation: for a partially filled level we add terms which correspond to the level as an empty one and as a filled one with coefficients $(1 - \{\nu\})$ and $\{\nu\}$ respectively. For example, if some level is half-filled then all the terms in the polarization function which correspond to the hopping to this level appear with the coefficient $1/2 = 1 - 1/2$, and the terms which correspond to hopping from this level also appear with the coefficient $1/2$. Thus, we do not take into account correlations in a partially filled LL.

For this work the polarization function was calculated approximately: we calculated only terms with $n, n' \leq n_{\mathrm{cutoff}} = 4$. We checked that the pseudopotentials in the region we are interested in differ from the pseudopotentials calculated with $n_{\mathrm{cutoff}} = 3$ by less than 2\%.

\subsubsection{\label{PopRevDet}Population reversal of Landau levels and level mixing}

It was discussed in sub-subsection~\ref{PopRevGen} that when the typical energy scale of the electron-electron interaction becomes larger than the difference of kinetic energies of two LLs from the same valley it is natural to expect the SLLA to break down. The numerical study in such cases is significantly hampered. Moreover, it is hardly probable to find a state similar to a single-level state in this regime of strong level mixing. Therefore, we would like to work only in the regime where the mixing of LLs is small. For this we demand the typical energy scale of the electron-electron interaction to be smaller than the kinetic energy distance to the closest LL from the same valley. The remnants of the level mixing can be incorporated then into the corrections to the SLLA which are discussed in the next sub-subsection.

One can use different quantities to define the typical energy scale of the electron-electron interaction. For example, one can use the interaction potential value at the magnetic length distance $V(l)$ or the zeroth pseudopotential at the LL one is interested in $V_{0}^{n, \xi, s}$. They typically differ by a factor of order of unity, which is not too important. We choose to use value of the zeroth pseudopotential as the typical interaction energy scale. Therefore, we restrict the region of consideration to those values of external parameters $U$, $B$, $\kappa$ for which $V_{0}^{n, \xi, s} \leq \Delta E$, where $\Delta E$ is the kinetic energy difference between the LL under consideration and the closest other LL to it.

There is a subtlety regarding this restriction. One can use the zeroth pseudopotential for the screened or for the bare Coulomb interaction potential. Using the Coulomb pseudopotential seems natural as it is the fundamental perturbation theory controlling parameter. However, for the weakened screened potential the Landau level mixing is smaller, and restricting the applicability region by the Coulomb interaction scale one excludes regions where our approach should still give reliable results. On the other hand, if the screening is so strong that the Coulomb energy scale significantly differs from the screened one, then the RPA approach we use for calculation of the screened potential may be not good enough, bringing in an error in the interaction potential. In section \ref{PossObsPfaffBLG} we restrict our region of consideration by the screened energy scale, however, for comparison we also show the region's boundary calculated for the unscreened Coulomb potential.

Apart from mixing of LLs from one valley, a similar process can take place for neighbouring levels from different valleys like $(2, \pm1)$ (see Fig.~\ref{fig:spectr}). This is due to Coulomb interaction on the lattice scale that can make electrons jump from one valley to another. This interaction is considered in more detail in \cite{Kharitonov_BLG_IntervalleyInteraction}. What is important for us is that the typical energy scale of this interaction is
\eq{\tilde{V}_{\mathrm{Coul}}(q)\left|_{q = \frac{4 \pi l}{3 a}}\right. = \frac{2 \pi e^2}{\kappa q l} \left|_{q = \frac{4 \pi l}{3 a}}\right. = \frac{3 e^2 a}{2 \kappa l^2},}
where $a \approx 0.25 \mathrm{nm}$ is the graphene lattice constant. We would like this interaction not to play a significant role. Therefore we restrict the parameters region by demanding that its energy scale is smaller than the distance between the LL under consideration and its closest neighbor from the different valley.\footnote{Unlike the case of level mixing in one valley, due to quasi-momentum conservation, mixing of the LLs from different valleys can happen only if both of them are filled with electrons at least partially. Naively, one would think that because of this argument the level $(2, +1)$ is not dangerous when we consider the $(2, -1)$ LL. However, the screening processes happen because of hopping of the electrons to higher LLs. Therefore, the $(2, +1)$ LL is "virtually" filled, to some extent. Thus, to be on the safe side, we still apply this restriction when we consider the $(2, -1)$ LL.}

Even if the valley index is a good quantum number (i.e. electron jumping between valleys is suppressed), the ground state might not be valley polarized. The reason for this can be the competition between the kinetic energy favoring a LL from one valley and the exchange energy favoring a LL from the opposite valley. If this happens, the two LLs, even not mixing, influence each other through the density-density interaction (since the electrons still repel each other). In such case the two levels from different valleys should be considered together just like in the case of level mixing. So by the same reasoning as in the case of mixing, we do not want to consider the system in the regime of two levels from different valleys partially filled.

Thus, we need to find the region of parameter space where such simultaneous filling does not occur~--- in order to use the SLLA there. Since the case of a partially filled level is hard to analyze, as a criterion we choose to demand that for integer fillings there should be no change of the filling order. I.e., the full energy (kinetic plus interaction) per electron of fully filled levels should put them in the same order as their kinetic energy. For example, if the kinetic energy of the $(2, -1)$ LL is less than the one of the $(2, +1)$ LL, then the full energy per electron in the fully filled $(2, -1)$ level should also be less than the full energy per electron in the fully filled $(2, +1)$ level.

For the fully filled level it is easy to calculate its interaction energy since there is only one state possible~--- the Slater determinant of all the level's orbitals. The interaction energy can be expressed through the pseudopotentials:
\begin{gather}
E_{\mathrm{inter.},\ N\ \mathrm{electrons}} = \frac{N(N-1)}{2} \text{Tr }\hat{V}\hat{\rho}_2,\\
E_{\mathrm{inter.}\ \mathrm{pp}} = \lim_{N \rightarrow \infty} \frac{E_{\mathrm{inter.},\ N \mathrm{electrons}}}{N} = 2 \sum_{m \in 2\mathbb{N}-1} V_{m},
\end{gather}
where $\hat{\rho}_2$ is the two-particle density matrix, $\mathrm{pp}$ in the subscript stands for "per particle".

This energy can be separated into the Hartree (density-density interaction) and the Fock (exchange) contributions:
\begin{equation}
\label{Hartree_energy_pp}
E_{\mathrm{Hartree}\ \mathrm{pp}} = \sum_{m = 0}^{\infty} V_{m} = \frac{1}{2 l^2} \int_{0}^{\infty} dr\ r V(r),
\end{equation}
\begin{equation}
E_{\mathrm{Fock}\ \mathrm{pp}} = \sum_{m = 0}^{\infty} (-1)^{m+1} V_{m} = \frac{1}{2 l^2} \int_{0}^{\infty} dr\ r V(r) g(r).
\end{equation}

The function $g(r)$ is related to two-particle and one-particle density matrices $\rho_2$ and $\rho_1$:
\eq{
\rho_1(x'|x) = \langle x'|\hat{\rho}_1| x \rangle,
}
\eq{
\rho_2(x'y'|xy) = \langle x'|\otimes\langle y'|\ \hat{\rho}_2\ | x \rangle \otimes |y\rangle,
}
\begin{multline}
\rho_2(x'y'|xy) ={}\\
 {}\left(\rho_1(x'|x)\rho_1(y'|y) - \rho_1(y'|x)\rho_1(x'|y)\right),
\end{multline}
\nmq{
g(r) = N^2 (\rho_2(r,0|r,0) - \rho_1(0|0)^2) =\\
 - N^2 \rho_1(0|r)\rho_1(r|0).
}

Note, that the Hartree energy can be expressed as an integral of the interaction potential, with the form of the integral independent of a Landau level. This is a consequence of the fact that the fully filled LL has constant density. We note that the integral for the Hartree energy is divergent at the upper limit for the Coulomb-like interaction potentials. However, only differences of the energies have physical meaning, thus we can calculate this integral with a certain regularization if we keep the regularization always the same.

The Fock contribution is, on the contrary, convergent, but it depends on the LL through the function $g(r)$, which characterizes short-range correlations.

If the interaction potential $V(r)$ is the same for two different Landau levels, their Hartree energies are identical. However, as the screening can be different for different LLs, their Hartree energies \eqref{Hartree_energy_pp} are different. One may think that this energy difference can contribute to population order reversal. However, one has to take into account the background positive charge (since the system is electrostatically neutral). The total electrostatic energy then, as we show in Appendix~\ref{PopRev_Hartree}, does not differ for differently screened potentials.

Therefore, the interaction energy difference comes from the Fock term only. For the non-relativistic levels and the Coulomb potential, the magnitude of the Fock energy decreases with the level number $n$, while having a negative sign. Thus, both the Coulomb interaction and the kinetic energy favour the natural LL population order. This is not the case in bilayer graphene.

Consider, for example, two levels $(2,-1)$ and $(2,+1)$ in BLG. The latter level has a greater kinetic energy. The wave functions in the $(2,+1)$ level are close to the wave functions of the non-relativistic $n = 0$ LL, while the wave functions in $(2,-1)$ are close to the ones in $n = 2$. Thus, the Fock energy prefers the $(2,+1)$ LL, while the kinetic energy prefers the $(2,-1)$ LL. Therefore, if the interaction is strong enough population reversal may happen.

While for the bare Coulomb interaction population reversal would happen in some regions of the parameter space, for the screened potentials we do not find such an effect for the $(2,-1)$ and $(2,+1)$ LLs. This makes emergence of valley-unpolarized states improbable.

In summary, in this sub-subsection we considered the effects of mixing of LLs from one valley, mixing of LLs from different valleys, and emergence of valley-unpolarized states. We find that the last one does not occur in realistic conditions. The other two effects restrict the applicability of the SLLA.

\subsubsection{\label{hoppingQM}Calculation of corrections to the SLLA pseudopotentials due to virtual hopping}

Suppose we are in the region where spin-/valley-unpolarized states do not emerge, and the LL mixing of the level under consideration with the levels from the same valley is small. Then the small mixing can be taken into account with the help of corrections to the electron-electron interaction within the SLLA. Those are the small corrections to the pseudopotentials. It is known that small corrections of the order of $5-10 \%$ to the pseudopotentials' values can lead to a significant change of the overlap with a trial state (see, e.g., Ref~\cite{DasS}). In this sub-subsection we present the formulae we use to compute such corrections.

Consider the two-particle problem. In the subsection \ref{2pBLG}, it was shown that the eigenstates of the two-particle problem within the SLLA are $|m, M\rangle$, with their energies being $V_{m}^{j}$, $j = (n, \xi, s)$. Let us denote $|m, M\rangle$ as $|m, M, j, j\rangle$ to emphasize that both of the electrons are in the LL $j$. Now we add to our consideration the closest unfilled LLs from the same valley (for small deviations from the naive SLLA those are the levels above); we introduce the following basis in the Hilbert space:
\eq{
\label{Eq_twoElectrState_DiffLLs_BLG_AntiSymm}
|m,M,j,j'\rangle =
\left\{
\begin{array}{c}
|m,M,j,j\rangle\rangle, \text{ if } j = j',\\
\frac{|m,M,j,j'\rangle\rangle + (-1)^{m+1} |m,M,j',j\rangle\rangle}{\sqrt{2}}, \text{ if } j \neq j',
\end{array}
\right.
}
where $m \in 2\mathbb{Z}_+ + 1$, $M \in \mathbb{Z}_+$, and
\nmq{
\label{Eq_twoElectrState_DiffLLs_BLG}
|m,M,j,j'\rangle\rangle=\frac{1}{\sqrt{2^{m+M}m!M!}} \times\\
 (\hat{b}_1^\dag-\hat{b}_2^\dag)^m(\hat{b}_1^\dag+\hat{b}_2^\dag)^M(\Psi_{n,-n}^{\xi s})_1(\Psi_{n',-n'}^{\xi s'})_2.
}
Note that the basis vectors \eqref{Eq_twoElectrState_DiffLLs_BLG_AntiSymm} are antisymmetric with respect to electron permutations, while the auxiliary vectors \eqref{Eq_twoElectrState_DiffLLs_BLG} are not.

The leading correction to the eigenstates' energies, which is due to the virtual hopping to the higher LLs from the same valley, is given by the second order perturbation theory:
\nmq{
\label{hopping_corr}
E_m^{j} =V_{m}^j -\\
 \sum_{(j_1, j_2 \leq j_1) \neq (j, j), m', M'} \frac{|\langle m,M,j,j|\hat{V}|m',M',j_1,j_2\rangle|^2}{E_{j_1}^{\mathrm{kin}} + E_{j_2}^{\mathrm{kin}} - 2 E_{j}^{\mathrm{kin}}}.
}

Since the interaction potential $V$ is a function of $r^2$, one can show that the only non-zero matrix elements of all the $\langle m,M,j,j|\hat{V}|m',M',j_1,j_2\rangle$ are
\nmq{
\langle m,M,j,j|\hat{V}|m + (n_1-n)+(n_2-n),M,j_1,j_2\rangle =\\
 \sqrt{2}^{1-\delta_{j_1j_2}} V_{m}^{j,j,j_1,j_2},
}
\nmq{
V_{m}^{j,j,j_1,j_2} =\\
 \langle\langle m,M,j,j|\hat{V}|m + (n_1-n)+(n_2-n),M,j_1,j_2\rangle\rangle,
}
where $\delta_{j_1j_2}$ is equal to $1$ when $j_1 = j_2$ and $0$ otherwise. We have used the fact that
\eq{
V_{m}^{j,j,j_1,j_2} = (-1)^{(n_1-n)+(n_2-n)} V_{m}^{j,j,j_2,j_1}.
}

The auxiliary matrix elements $V_{m}^{j,j,j_1,j_2}$ can be expressed through non-relativistic matrix elements, similarly to how the pseudopotentials in BLG are expressed via the non-relativistic pseudopotentials:
\nmq{
|m,M,n,n'\rangle\rangle = \frac{1}{\sqrt{2^{m+M}m!M!}} \times\\
 (\hat{b}_1^\dag-\hat{b}_2^\dag)^m(\hat{b}_1^\dag+\hat{b}_2^\dag)^M(\psi_{n,-n})_1(\psi_{n',-n'})_2,
}
\nmq{
V_{m}^{n,n',n_1,n_2} =\\
 \langle\langle m,M,n,n'|\hat{V}|m + (n_1-n)+(n_2-n'),M,n_1,n_2\rangle\rangle,
}
\eq{
V_{m}^{n,n,n_1,n_2} = (-1)^{(n_1-n)+(n_2-n)} V_{m}^{n,n,n_2,n_1},
}
\nmq{
V_{m}^{j,j,j_1,j_2} = V_m^{n,n,n_1,n_2}\overline{A_{n}^{\xi s}}^2 A_{n_1}^{\xi s_1}A_{n_2}^{\xi s_2} +\\
 V_m^{n-2,n,n_1-2,n_2}\overline{B_{n}^{\xi s}} \overline{A_{n}^{\xi s}} B_{n_1}^{\xi s_1} A_{n_2}^{\xi s_2}+...
}

The non-relativistic matrix elements can be expressed through the Fourier transform of the pseudopotential in a form quite similar to the expression for the pseudopotentials \eqref{pseudopotFourier}:
\begin{multline}
V_m^{n,n',n_1,n_2} = \int_{0}^{\infty} \tilde{V}(q) F_{m,m + (n_1-n)+(n_2-n')}(\bar{x}\sqrt{2})\times{}\\
 {}F_{n,n_1}(x) F_{n',n_2}(-x) e^{-q^2} \frac{q dq}{2 \pi},
\end{multline}
\begin{multline}
F_{i_1,i_2}(x)|_{i_1 \geq i_2} ={}\\
 {}\sum_{k = 0}^{i_2} \frac{\sqrt{i_1! i_2!}}{k! (k+i_1-i_2)! (i_2 - k)!} \left(-\frac{|x|^2}{2}\right)^k \left(\frac{i x}{\sqrt{2}}\right)^{i_1-i_2} ={}\\
  {} \sqrt{\frac{i_2!}{i_1!}} \left(\frac{i x}{\sqrt{2}}\right)^{i_1-i_2} L_{i_2}^{(i_1-i_2)}(|x^2|/2),
\end{multline}
\begin{equation}
F_{i_1,i_2}(x)|_{i_1 \leq i_2} = F_{i_2,i_1}(\bar{x}),
\end{equation}
$x = q_x + i q_y$, and $L_{n}^{(\alpha)}$ are generalized Laguerre polynomials. The derivation is very similar to the derivation of formula \eqref{pseudopotFourier} presented in Appendix~\ref{FourierFormula}. Since $F_{n,n}(x) = L_{n}(|x|^2/2)$, for $n = n_1$, $n' = n_2$ the formula \eqref{pseudopotFourier} is restored, as it should be because by definition $V_m^{(n_1,n_2)} = V_m^{n_1,n_2,n_1,n_2}$.

There is a subtlety regarding what interaction potential $\hat{V}$ should be used in the formulas above. It is tempting to use the screened interaction potential \eqref{screening} to calculate virtual hopping corrections. However, as we show in Appendix~\ref{peculiDiag}, this would exceed the accuracy of the perturbative calculation. Therefore, we use the unscreened Coulomb interaction potential in calculating the virtual hopping corrections to the SLLA pseudopotentials.

\subsection{Summary of the section}

In this section we discuss the effects which restrict the applicability of the SLLA in BLG. We also discuss the conditions under which it is sufficient to introduce corrections to the SLLA to restore the theory applicability. We calculate these corrections in the second subsection.

\section{\label{PossObsPfaffBLG}Possibility to observe the Moore-Read state in bilayer graphene}

In order to investigate the role of the effects discussed above on the stability of FQHE states we focus on the Moore-Read Pfaffian. Our choice is motivated by the following considerations. Firstly, this state is particularly sensitive to the details of the interaction so it is a good illustration for our analysis. Secondly, the stability of this state in BLG was investigated in Refs.~\cite{Apalkov2011, Abanin2011} in the SLLA approximation but without these effects taken into account, so we can compare the phase diagrams. Thirdly, the Moore-Read state itself is an important state because it is an example of the non-abelian topological fluid.

The tunable parameters are the magnetic field $B$, the electric field which determines the mini-gap parameter $U$ and the effective dielectric constant $\kappa$\footnote{The dielectric constant sensed by BLG is $\kappa = (\kappa_1+\kappa_2)/2$, where $\kappa_1$ and $\kappa_2$ are the dielectric constants of the environment below and above the sheet of BLG.} which controls the deviation from the naive SLLA (which is exact for $\kappa \rightarrow \infty$). We can also choose the half-filled LL number. Here we will concentrate only on the two levels: $(1, -1)$ and $(2, -1)$. The $(1, -1)$ level wave function is constructed from the nonrelativistic $n = 0$ and $n = 1$ LL wave functions, the $(2, -1)$ level wave function is constructed from the nonrelativistic $n = 0,1,2$ LL wave functions. In both cases for the bare Coulomb interaction one can tune the pseudopotentials close to their values at the nonrelativistic $n = 1$ LL, where the $5/2$ state is observed in GaAs\footnote{Though the $(2, +1)$ level wave function is also constructed from the nonrelativistic $n = 0,1,2$ LL wave functions, the numerics shows that the high overlap with the Moore-Read state is not achieved here for the bare Coulomb interaction in contrast to the $(1, -1)$ and the $(2, -1)$ LLs.}.

The tuning mechanisms are, however, different for the two levels. Amplitudes of the wave function \eqref{wavefunction} in the $(1, -1)$ LL show little dependence on $U$ so the main control parameter is $B$\footnote{In the low-energy two-band model \cite{MCF} (which corresponds to the $\gamma_1 \rightarrow \infty$ limit) such tuning is impossible because the amplitude $A_{1}^{-1,+1}$ is identically equal to $1$ with other amplitudes being zero.}. In contrast, the amplitudes of the wave function in the $(2, -1)$ LL mainly depend on one parameter which is the $U/\hbar \omega_c$ ratio, so both $B$ and $U$ can be used for tuning.

The main factors determining deviation from the naive SLLA for the two levels are the polarization and virtual hopping to the nearby levels. For the $(1, -1)$ LL this is hopping to the $(0, -1)$ and the $(2, -1)$ LLs, while for the $(2, -1)$ LL the important hopping is to the $(3, -1)$ LL. In addition to this, for the $(2, -1)$ LL, it is important to consider effects of mixing with the $(0, -1)$ and $(2, +1)$ LLs. The latter are important factors restricting the region of applicability of perturbative analysis, however, when suppressed they do not lead to a renormalization of the intra LL interaction.

\begin{figure}[tbp]
  \centering
  \includegraphics[width=.96\columnwidth]{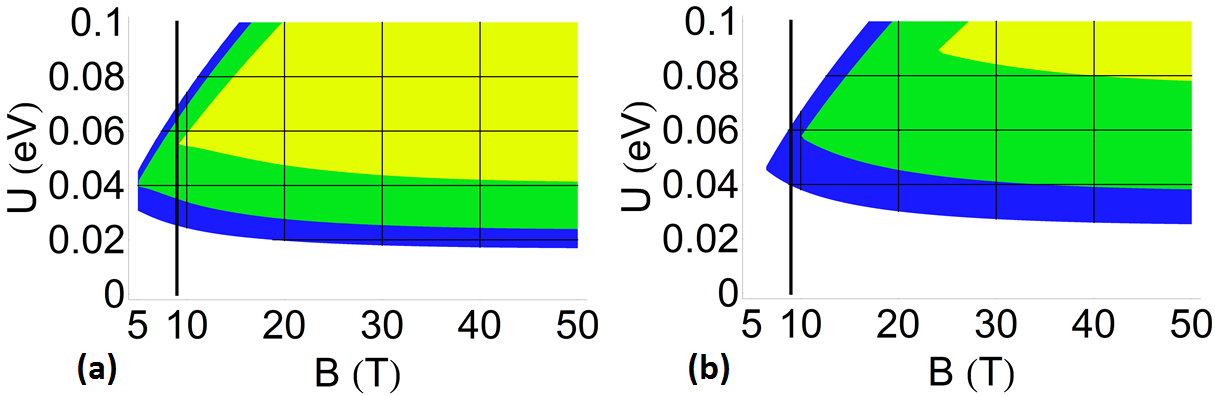}
  \caption{\textbf{The region of the applicability of perturbative analysis for fixed values of $\kappa = 5$ (yellow region), $10$ (green and yellow regions) and $15$ (blue, green, and yellow regions) for the $(1, -1)$ LL.} The size of the region increases with increasing $\kappa$. For $(a)$ the typical interaction energy scale is taken to be the zeroth pseudopotential of the screened interaction potential, for $(b)$~--- of the bare Coulomb potential. The thick black line shows where the maximum overlap with the Moore-Read Pfaffian state for the bare Coulomb interaction is achieved. (Color online).}
  \label{fig:zones1}
\end{figure}

Figures~\ref{fig:zones1}a and \ref{fig:zones1}b show the regions of the applicability of perturbative analysis for different values of $\kappa$ for the $(1, -1)$ LL.\footnote{\label{AboutMistakes}Due to some errors in calculation of the region of applicability, in the original result-reporting paper \cite{SniCheSim} the form of the region is not entirely correct. This applies to both the $(1, -1)$ LL and the $(2, -1)$ LL case, which is considered next.} For Fig.~\ref{fig:zones1}a the typical interaction energy scale, which determines the significance of level mixing, is estimated with the help of the screened potential ("type S estimate"), while for Fig.~\ref{fig:zones1}b~--- with the help of the bare Coulomb potential ("type C estimate"). The regions are bounded from above by the condition of small hopping to the $(2, -1)$ LL, the lower bound is due to the condition of small hopping to the $(0, -1)$ LL. At small enough magnetic fields at least one of the conditions is violated at all values of $U$. The thick black line shows where the maximum overlap\footnote{By overlap we mean scalar product's absolute value squared.} with the Moore-Read Pfaffian for the bare Coulomb interaction (which is about $0.95$, compare with non-relativistic $n = 1$ level overlap of $0.7$) is achieved. One can see that for small dielectric constants this line lies outside the region of validity of perturbative analysis, however for large enough $\kappa$ they intersect near $U = 50\ \text{meV}$.

\begin{figure}[tbp]
  \centering
  \includegraphics[width=.96\columnwidth]{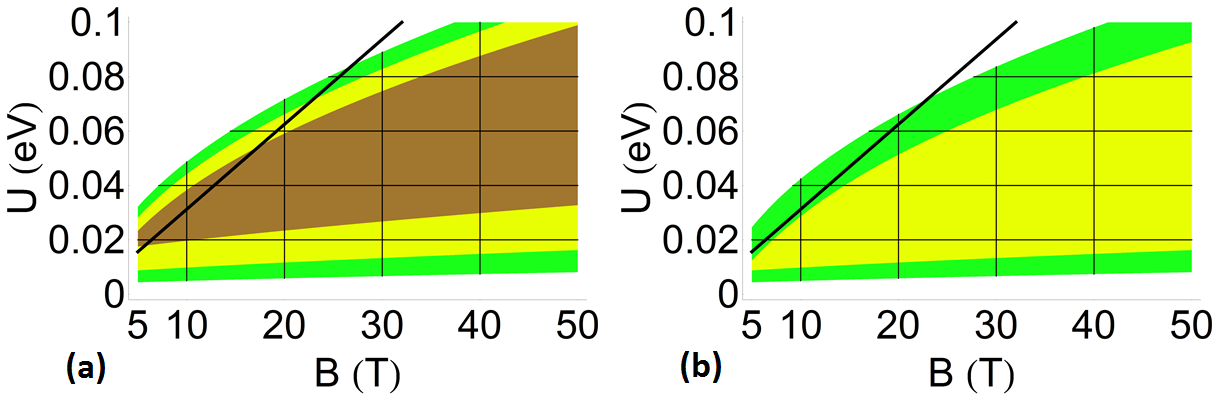}
  \caption{\textbf{The region of the applicability of perturbative analysis for fixed values of $\kappa = 2.5$ (brown region), $5$ (yellow and brown regions) and $10$ (green, yellow, and brown regions) for the $(2, -1)$ LL.} The size of the region increases with increasing $\kappa$. For $(a)$ the typical interaction energy scale is taken to be the zeroth pseudopotential of the screened interaction potential, for $(b)$~--- of the bare Coulomb potential. At $\kappa = 2.5$ in $(b)$ the condition of smallness of level mixing is not satisfied anywhere within the range of external parameters shown. The thick black line shows where the maximum overlap with the Moore-Read Pfaffian state for the bare Coulomb interaction is achieved. (Color online).}
  \label{fig:zones2}
\end{figure}

Figures~\ref{fig:zones2}a and \ref{fig:zones2}b show the regions of the applicability of perturbative analysis for different values of $\kappa$ for the $(2, -1)$ LL. For Fig.~\ref{fig:zones2}a the type S estimate is used, while for Fig.~\ref{fig:zones2}b the type C estimate is used. The regions are bounded from above by the condition of small mixing with the $(0, -1)$ LL, the lower bound is due to the condition of small mixing with the $(2, +1)$ LL. The thick black line shows where the maximum overlap with the Moore-Read Pfaffian for the bare Coulomb interaction (which is about $0.92$, compare with non-relativistic $n = 1$ level overlap of $0.7$) is achieved. One can see that for small dielectric constants this line lies outside the region of validity of perturbative analysis, however for large enough $\kappa$ they intersect near $U = 30\ \text{meV}$.

\begin{figure}[tbp]
  \centering
  \includegraphics[width=.96\columnwidth]{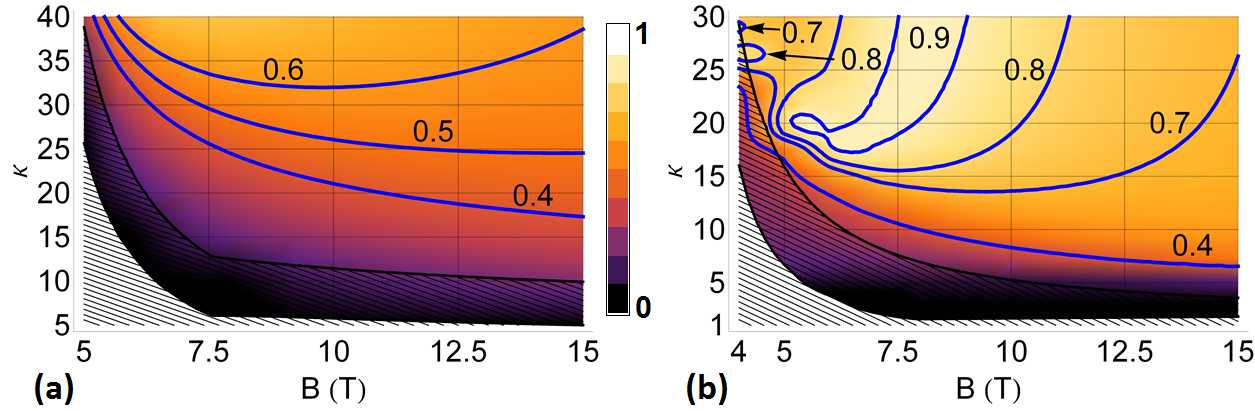}
  \caption{\textbf{Color plot of the overlap of the ground state with the Moore-Read Pfaffian for $12$ particles as a function of the magnetic field $B$ and the dielectric constant $\kappa$.} $(a)$~-- for the $(1, -1)$ LL at $U = 50\ \text{meV}$, $(b)$ -- for the $(2, -1)$ LL at $U = 30\ \text{meV}$. Contours show the lines of constant overlap. The region where perturbative analysis is not applicable according to the type C estimate is hatched. Data is not shown beyond the region where perturbative analysis is applicable according to the type S estimate. (Color online).}
  \label{fig:overlap}
\end{figure}

Figures \ref{fig:overlap}a and \ref{fig:overlap}b show the dependence of the overlap of the exact ground state of the system with the Pfaffian on the magnetic field and the dielectric constant at $U = 50\ \text{meV}$ for the $(1, -1)$ LL and at $U = 30\ \text{meV}$ for the $(2, -1)$ LL respectively.\footnote{Due to some mistakes in calculation of the screened potential, in the original result-reporting paper \cite{SniCheSim} the overlaps for the $(1, -1)$ LL behave somewhat differently. Here we corrected the mistakes. As to the $(2, -1)$ LL, here we consider another value of $U$ than in the paper \cite{SniCheSim}, see footnote \ref{AboutMistakes}.}  We do not show the data in the region where the perturbative analysis is not applicable according to type S estimate. The region of inapplicability of our theory according to the type C estimate is hatched. As one can see, for the $(1, -1)$ level, the highest overlap achieved at dielectric constants as high as $\kappa = 40$ is about $0.7$ despite having overlaps up to $0.95$ for the bare Coulomb interaction (which corresponds to the $\kappa \rightarrow \infty$ limit). For the $(2, -1)$ level a high overlap up to $0.92$ is achieved at $\kappa \gtrsim 20$.

The authors of \cite{Apalkov2011} found that in the $(1, -1)$ LL, high overlap is achieved in the region near $B = 10\ \text{T}$. We find that the region of high overlap is situated there for $\kappa \gtrsim 60$. However, for realistic values of $\kappa$ the effects of level mixing become significant which makes observation of the Moore-Read state unlikely, especially for $\kappa \lesssim 10$.\footnote{It is interesting to note that recently a $\nu = 1/2$ FQHE state has been observed in suspended bilayer graphene \cite{MorpurgoFalko_HalfFilling_BLG_Observation} for $U \approx 0$. The parameters of the experiment lie far outside the region of applicability of our methodology. However, the study \cite{PapicAbanin_HalfFilling_BLG_Pfaffian}, which does take into account the strong mixing between the quasidegenerate $(1, -1)$ and $(0, -1)$ LLs beyond perturbation theory, claims that the Moore-Read Pfaffian is a likely candidate to explain the observed $\nu = 1/2$ FQHE.} The $(2, -1)$ LL was also considered in \cite{Apalkov2011}, where it was concluded that the maximal overlap with the Moore-Read Pfaffian on this level is less than $0.6$. Our results do not support this conclusion (even for the bare Coulomb interaction).

Thus, our results show that the $(2, -1)$ LL is a better candidate to observe the Moore-Read Pfaffian state in BLG than the $(1, -1)$ LL. However, even in the case of the $(2, -1)$ LL to tune into the regime of high overlap with the state one needs a dielectric constant about $\kappa = 20$. This is much higher than the usual $\kappa \approx 2.5$ for graphene on $\rm{SiO_2}$ or hexagonal boron nitride (hBN) substrate.  Even on $\rm{HfO_2}$ substrate \cite{Hafnium} $\kappa$ is around $12.5$. A possible way to achieve higher dielectric constants is to use substrates on the both sides of the BLG sheet. In that case the effective dielectric constant $\kappa$ is equal to the substrate dielectric constant. Therefore, using $\rm{HfO_2}$ one can get $\kappa = 25$.\footnote{Such configurations with hBN as a substrate have recently started being explored experimentally from the perspective of FQHE: see Ref.~\cite{TunableFQHE_BLG_Observation}.}

The region of high overlap intersects this value of dielectric constant for the $(2, -1)$ LL. For example, at $U = 30\ \text{meV}$, $B = 8\ \text{T}$ the overlap of the exact ground state and the Moore-Read state is about $0.92$, with the gap to the first excited state being about $1.6\ \text{K}$. With increasing magnetic field the gap monotonically increases to the values of around $5\ \text{K}$ at $B = 15\ \text{T}$. At the same time the overlap decreases to around $0.7$ which is still fairly large. Similarly, with decreasing the dielectric constant down to $\kappa = 10$ (for which one still needs $\rm{HfO_2}$ substrate but only on one side of the BLG sheet) the gap monotonically increases to the values of around $3\ \text{K}$, with the overlap decreasing to $0.4$ which is not too small.\footnote{In our simulations the Hilbert space is about 16000-dimensional. Therefore, two random vectors would have the average overlap about 1/16000.} This result, obtained for a finite number of particles, suggests that the system may still be in the same topological phase at higher magnetic fields and lower dielectric constants.

However, the experimental observation of the Moore-Read Pfaffian in BLG in this way in the near future is unlikely. The main problem for observation of the fractional QHE in graphene (and BLG as well) is the too high disorder in the samples caused by the disorder in the substrate (evidently, it is impossible to observe FQHE when the typical height of the disorder potential is greater than the gap to the first excited state). For example, FQHE with the filling factor $\nu = 1/3$ has been observed in suspended single-layer graphene \cite{Bolotin2009} and in graphene on hBN substrate \cite{FQHE_graphene_hBN_obsrevation} but not on a substrate with a different lattice structure (which is the case for $\rm{HfO_2}$ needed to achieve the high dielectric constant).

The reader can find some additional data on the behaviour of gaps and overlaps in Appendix~\ref{PfaffSupplMat}.

One important thing to note in the context of modern experiments is that the possibility to tune the mini-gap parameter $U$ is achieved through placing two metallic gates on the both sides of the BLG sheet, very close to it \cite{TunableFQHE_BLG_Observation}. Therefore, one can expect that the screening of the Coulomb potential will be due to not just the internal BLG dynamics but also due to the metallic gates. We have not taken this effect into account in our analysis.

\section*{Conclusions}

We analyze the influence of inter-Landau level transitions (vacuum polarization, virtual hopping) on the phase diagram of the FQHE states. We find that the SLLA can only be used under quite stringent conditions, and corrections to the SLLA should be taken into account. A method for taking the corrections into account by means of perturbation theory is developed. However, the region of applicability of the SLLA with our corrections is also quite restricted.

With the help of the developed method we study the possibility to observe the Moore-Read state in bilayer graphene. We find that the mentioned effects indeed lead to a substantial modification of the phase diagram. However, the external parameters needed to tune into the regime favouring the Moore-Read state are, in principle, achievable, for BLG surrounded with $\rm{HfO_2}$.

We also developed a set of programs in Wolfram Mathematica which implement our methodology to study the case of Moore-Read Pfaffian state. To obtain the programs please contact K.S.

\section*{Acknowledgements}

We thank Prof. V. Fal'ko for useful discussions. We thank the authors of DiagHam code for numerical diagonalization, especially N. Regnault. We also thank the authors of the Jaxodraw program, which was used to draw the Feynman diagrams.

\appendix

\section{\label{FourierFormula}Derivation of the expression for pseudopotentials in terms of the Fourier transform of the interaction potential}

In this appendix we derive an integral representation for the pseudopotential
\nmq{
\label{defPseudopot}
V_m^{(n_1,n_2)}=\langle n_1,n_2,m,M| \hat{V} |n_1,n_2,m,M\rangle =\\ \langle n_1,n_2,m,0| \hat{V} |n_1,n_2,m,0\rangle,
}
where the two-electron states
\nmq{
|n_1,n_2,m,M\rangle=\frac{1}{\sqrt{2^{m+M}m!M!}} \times\\
 (\hat{b}_1^\dag-\hat{b}_2^\dag)^m (\hat{b}_1^\dag+\hat{b}_2^\dag)^M (\psi_{n_1,-n_1})_1(\psi_{n_2,-n_2})_2
}
have been defined in section \ref{SubSect_TwoElectrProblem_NonRel}, Eq.~\eqref{Eq_twoElectrState_DiffLLs}.

The direct Fourier transform for the interaction potential is defined in \eqref{potFourier}. The inverse Fourier transform looks like
\begin{equation}
\label{invFourier}
V(r) = \int \frac{d^2q}{(2\pi)^2} \tilde{V}(q) e^{i \mathbf{q} \mathbf{r}/l}.
\end{equation}

We can rewrite the definition \eqref{defPseudopot} as follows:
\begin{multline}
V_m^{(n_1,n_2)}=\langle n_1,n_2,m,M| \hat{V} |n_1,n_2,m,M\rangle ={}\\
{} \int \frac{d^2q}{(2\pi)^2} \tilde{V}(q) \langle n_1,n_2,m,M| e^{i \mathbf{q} \hat{\mathbf{r}}/l} |n_1,n_2,m,M\rangle.
\end{multline}

Introducing $x = q_{x} + i q_{y}$, we can write $\mathbf{q} \mathbf{r}/l = (x \bar{w} + \bar{x} w)/2$. $w, \bar{w}$ can be expressed through the $\hat{a}, \hat{a}^\dag, \hat{b}, \hat{b}^\dag$ operators:
\begin{equation}
w = \sqrt{2}(\hat{a} + \hat{b}^\dag),\ \bar{w} = \sqrt{2}(\hat{a}^\dag + \hat{b}).
\end{equation}

Thus, the matrix element can factorized into a product of three different matrix elements:
\begin{equation}
\langle n_1,n_2,m,M| e^{i \mathbf{q} \hat{\mathbf{r}}/l} |n_1,n_2,m,M\rangle = \mathcal{M}_{n_1}\mathcal{M}_{n_2}\mathcal{M}_{m},
\end{equation}
where
\begin{eqnarray}
\label{Eq_Mn1}
\mathcal{M}_{n_1} &=& \langle n_1|e^{\frac{i}{\sqrt{2}}(x \hat{a}_1^\dag+\bar{x} \hat{a}_1)}|n_1\rangle,\\
|n_1\rangle &=& \frac{1}{\sqrt{n_1!}} (\hat{a}_1^\dag)^{n_1}|\Omega\rangle,\\
\hat{a}_1|\Omega\rangle &=& 0,
\end{eqnarray}
\begin{eqnarray}
\label{Eq_Mn2}
\mathcal{M}_{n_2} &=& \langle n_2|e^{-\frac{i}{\sqrt{2}}(x \hat{a}_2^\dag+\bar{x} \hat{a}_2)}|n_2\rangle,\\
|n_2\rangle &=& \frac{1}{\sqrt{n_2!}} (\hat{a}_2^\dag)^{n_2}|\Omega\rangle,\\
\hat{a}_2|\Omega\rangle &=& 0,
\end{eqnarray}
\begin{eqnarray}
\label{Eq_Mm}
\mathcal{M}_{m} &=& \langle m|e^{\frac{i}{\sqrt{2}}(x (\hat{b}_1-\hat{b}_2)+\bar{x} (\hat{b}_1^\dag-\hat{b}_2^\dag))}|m\rangle,\\
|m\rangle &=& \frac{1}{\sqrt{2^m m!}} (\hat{b}_1^\dag-\hat{b}_2^\dag)^{m}|\Omega\rangle,\\
(\hat{b}_1-\hat{b}_2)|\Omega\rangle &=& 0.
\end{eqnarray}

We note that the expressions \eqref{Eq_Mn1}, \eqref{Eq_Mn2}, \eqref{Eq_Mm} are similar and can be calculated using the same technique. We only give the details of the calculation of $\mathcal{M}_{n_1}$. By virtue of the Baker-Campbell-Hausdorff identity,
\begin{equation}
e^{\frac{i}{\sqrt{2}}(x \hat{a}_1^\dag+\bar{x} \hat{a}_1)} = e^{\frac{i}{\sqrt{2}} x \hat{a}_1^\dag} e^{\frac{i}{\sqrt{2}} \bar{x} \hat{a}_1}e^{-|x|^2/4}.
\end{equation}
It follows that
\begin{multline}
\mathcal{M}_{n_1} = \langle n_1|e^{\frac{i}{\sqrt{2}}(x \hat{a}_1^\dag+\bar{x} \hat{a}_1)}|n_1\rangle = {}\\
{}e^{-|x|^2/4} \langle n_1|e^{\frac{i}{\sqrt{2}} x \hat{a}_1^\dag} e^{\frac{i}{\sqrt{2}} \bar{x} \hat{a}_1}|n_1\rangle = {}\\
{}e^{-|x|^2/4} \sum_{n = 0}^{\infty} \langle n_1|e^{\frac{i}{\sqrt{2}} x \hat{a}_1^\dag} |n\rangle\langle n| e^{\frac{i}{\sqrt{2}} \bar{x} \hat{a}_1}|n_1\rangle,
\end{multline}
where we used the resolution of identity $\hat{1} = \sum_{n = 0}^{\infty}|n\rangle\langle n|$.

After a calculation of the expression
\begin{multline}
\langle n| e^{\frac{i}{\sqrt{2}} \bar{x} \hat{a}_1}|n_1\rangle = \sum_{k = 0}^\infty \frac{1}{k!} \left(\frac{i}{\sqrt{2}} \bar{x}\right)^k \langle n|\hat{a}_1^k|n_1\rangle ={}\\
{}= \sum_{k = 0}^\infty \frac{1}{k!} \left(\frac{i}{\sqrt{2}} \bar{x}\right)^k \delta_{k,n_1-n} \sqrt{\frac{n_1!}{n!}} ={}\\
{} \frac{1}{(n_1-n)!} \left(\frac{i}{\sqrt{2}} \bar{x}\right)^{n_1-n} \sqrt{\frac{n_1!}{n!}},
\end{multline}
and its complex conjugate, we get
\begin{multline}
\mathcal{M}_{n_1} = e^{-|x|^2/4} \sum_{n = 0}^{n_1} \left(-\frac{|x|^2}{2}\right)^{n_1-n} \frac{n_1!}{n! (n_1-n)!^2} ={}\\
{} |k=n_1-n| ={}\\
{} e^{-|x|^2/4} \sum_{k = 0}^{n_1} \left(-\frac{|x|^2}{2}\right)^{k} \frac{n_1!}{(n_1-k)! (k)!^2} ={}\\
{} e^{-|x|^2/4} \sum_{k = 0}^{n_1} \left(-\frac{|x|^2}{2}\right)^{k} \frac{1}{k!} C_{n_1}^{k} = e^{-|x|^2/4} L_{n_1}(|x^2|/2),
\end{multline}
where $L_{n_1}$ is the Laguerre polynomial.

Calculation of $\mathcal{M}_{n_2}$, $\mathcal{M}_{m}$ is done the same way. Gathering all the three matrix elements together we finally get formula \eqref{pseudopotFourier}:
\begin{multline}
V_m^{(n_1,n_2)} = {}\\
 {}\int_{0}^{\infty} \tilde{V}(q) L_{m}(q^2) L_{n_1}(q^2/2) L_{n_2}(q^2/2) e^{-q^2} \frac{q dq}{2 \pi}.
\end{multline}

\section{\label{PopRev_Hartree}Full Hartree energy of a Landau level}

The electrostatic (Hartree) energy per electron of a fully filled Landau level can be expressed as
\eq{
E_{\mathrm{Hartree}\ \mathrm{pp}} = \frac{1}{2 N} \int d^{2}x\ d^{2}y \rho{(\mathbf{x})}\rho{(\mathbf{y})} V(|\mathbf{x} - \mathbf{y}|),
}
where $N$ is the number of electrons in the LL, $\mathbf{x}$ and $\mathbf{y}$ are the coordinate vectors of points in the plane, $V(r)$ is the electron-electron interaction potential, and $\rho{(\mathbf{x})}$ is the electron density, integration is done over the whole plane\footnote{Over the whole sample of area $S \rightarrow \infty$ with $N$ electrons in the LL so that $S/N = 2 \pi l^2$. $l$ is the magnetic length.}. Using the fact that for a fully filled LL $\rho{(\mathbf{x})} = 1/(2 \pi l^2) = \rho$ and changing integration variables to $\mathbf{R} = (\mathbf{x} + \mathbf{y})/2$ and $\mathbf{r} = \mathbf{x} - \mathbf{y}$, we get
\eq{
E_{\mathrm{Hartree}\ \mathrm{pp}} = \frac{\rho^2}{2 N} \int d^{2}R\ d^{2}r V(r),
}
where $r = |\mathbf{r}|$. $\int d^{2}R$ is equal to the sample area $S$. $\rho \times S = N$, therefore,
\nmq{
E_{\mathrm{Hartree}\ \mathrm{pp}} = \frac{\rho}{2} \int d^{2}r V(r) =\\
 \frac{1}{4 \pi l^2} 2 \pi \int dr V(r) = \frac{1}{2 l^2} \int_{0}^{\infty} dr\ r V(r),
}
in agreement with Eq.~\eqref{Hartree_energy_pp}.

This derivation is valid for the potentials that decay sufficiently fast as $r \rightarrow \infty$, and the Coulomb interaction does not satisfy this condition. However, only differences of the energies have physical meaning. If we consider two energies for differently screened interaction potentials \eqref{screening}:
\nmq{
E_{\mathrm{Hartree}\ \mathrm{pp}}^{(1)} - E_{\mathrm{Hartree}\ \mathrm{pp}}^{(2)} =\\
 \frac{1}{2 l^2} \int_{0}^{\infty} dr\ r \left(V^{(1)}(r) - V^{(2)}(r)\right),
}
the integral of the difference is convergent since the Coulomb part cancels out and what remains decays sufficiently fast at large distances. Therefore, we can express such difference through the interaction potential Fourier transforms $\tilde{V}(q)$ defined in Eq.~\eqref{potFourier}:
\nmq{
E_{\mathrm{Hartree}\ \mathrm{pp}}^{(1)} - E_{\mathrm{Hartree}\ \mathrm{pp}}^{(2)} =\\
 \frac{1}{4\pi} \lim_{q \rightarrow 0} \left(\tilde{V}^{(1)}(q)-\tilde{V}^{(2)}(q)\right).
}

Recalling \eqref{screening} and taking into account that $\Pi(q) = \Pi(q, \omega = 0) \sim \text{const}\times q^2 + O(q^4)$ for $q \rightarrow 0$, one gets
\nmq{
E_{\mathrm{Hartree}\ \mathrm{pp}}^{(1)} - E_{\mathrm{Hartree}\ \mathrm{pp}}^{(2)} =\\
 \frac{\pi e^4}{\kappa^2} \lim_{q \rightarrow 0} \left(\frac{\Pi^{(2)}(q)-\Pi^{(1)}(q)}{q^2}\right).
}

Based on this consideration, naively one might think that Hartree energies contribute to the population order reversal effect and that the contribution is characterized by the quantity $\lim_{q \rightarrow 0} \Pi(q)/q^2$. However, this consideration does not take into account the electrostatic energy of interaction with the compensating positive charge (as the system is electrically neutral as a whole). Although, the screening processes happen at the BLG plane, they influence the interaction potential of charges outside of the plane.

Consider a simple model: the compensating positive charge is evenly distributed in the plane at a distance $d$ from the BLG plane; for simplicity, all the system is in the environment with the dielectric constant $\kappa = 1$. Suppose the screening processes happen only in the BLG sheet. Then one can calculate the potential of interaction of the charges situated in different points of space. We are going to do that now.

\begin{widetext}
We need several definitions: Coulomb electrostatic potential $\varphi\left(r = \sqrt{x^2 + y^2 + z^2}\right) = 1/r$, the spatial polarization function with screening happening only in the $z = 0$ plane $\Pi(\mathbf{r}) = \Pi(x,y,z) = \Pi(x,y) \delta(z)$\footnote{$\delta(z)$ is the Dirac delta-function.}, and their Fourier transforms w.r.t. plane coordinates $x, y$ defined similarly to Eq.~\eqref{potFourier} and to Eq.~\eqref{PolFunc_Fourier} respectively: $\tilde{\varphi}(q = |\mathbf{q}|, z) = 2 \pi \exp{(-q |z|/l)}/(q l)$ and $\Pi(q,z) = \Pi(q) \delta(z)$.

Then the interaction potential of two charges $e_1$ and $e_2$ placed in $\mathbf{r}$ and $\mathbf{r}'$ is
\begin{multline}
V_{\mathrm{scr}}(\mathbf{r}, \mathbf{r}') = e_1 e_2 \varphi_{\mathrm{scr}}(|\mathbf{x} - \mathbf{y}|, z, z') =\\
 e_1 e_2 \varphi(\mathbf{r}, \mathbf{r}') -\\
  - e_1 e_2 e^2 \int d^{3}x d^{3}y \varphi(|\mathbf{r} - \mathbf{x}|)\Pi(\mathbf{x} - \mathbf{y})\varphi(|\mathbf{y} - \mathbf{r}'|) +\\
   + e_1 e_2 e^4 \int d^{3}x d^{3}y d^{3}z d^{3}w \varphi(|\mathbf{r} - \mathbf{x}|)\Pi(\mathbf{x} - \mathbf{y})\varphi(|\mathbf{y} - \mathbf{z}|)\Pi(\mathbf{z} - \mathbf{w})\varphi(|\mathbf{w} - \mathbf{r}'|) - ...
\end{multline}
Here $e$ is the elementary charge~--- the absolute value of the charge of the electrons participating in the screening processes.

One can show then that the in-plane Fourier transform of the screened interaction potential of the two charges has the form
\begin{multline}
\tilde{V}_{\mathrm{scr}}(q, z, z', e_1, e_2) = e_1 e_2 \tilde{\varphi}_{\mathrm{scr}}(q, z, z') =
 \\ e_1 e_2 \tilde{\varphi}(q, z-z') - e_1 e_2 \tilde{\varphi}(q, z) \frac{e^2 l^2 \Pi(q)}{1 + e^2 l^2 \tilde{\varphi}(q, 0) \Pi(q)} \tilde{\varphi}(q, -z') = \\
  e_1 e_2\Bigg(\frac{2 \pi \exp{\left(-\frac{q |z-z'|}{l}\right)}}{q l} - \frac{2 \pi \exp{\left(-\frac{q |z|}{l}\right)}}{q l} \frac{e^2 l^2 \Pi(q)}{1 + e^2 l^2 \frac{2 \pi}{q l} \Pi(q)}\frac{2 \pi \exp{\left(-\frac{q |z'|}{l}\right)}}{q l}\Bigg).
\end{multline}
For $e_1 = e_2 = e$ and $z = z' = 0$ one restores the expression \eqref{potFourier}, as it should be.

The electrostatic energy of interaction of all the electrons in the fully filled LL and the charge-compensating background is equal to
\begin{multline}
E_{\mathrm{Hartree}} = e^2 \int d^{2}x\ d^{2}y \rho{(\mathbf{x})}\rho{(\mathbf{y})} \Bigg(\frac{1}{2} \varphi_{\mathrm{scr}}(|\mathbf{x} - \mathbf{y}|, z=z'=0) +\\
 +\frac{1}{2} \varphi_{\mathrm{scr}}(|\mathbf{x} - \mathbf{y}|, z=z'=d) - \varphi_{\mathrm{scr}}(|\mathbf{x} - \mathbf{y}|, z=d, z'=0) \Bigg).
\end{multline}
Performing the same operations as in the beginning of the current appendix for the Hartree energy of electrons in the LL only, we get:
\begin{multline}
E_{\mathrm{Hartree}} = \frac{e^2 N}{2 \pi} \lim_{q \rightarrow 0} \Bigg(\frac{1}{2} \tilde{\varphi}_{\mathrm{scr}}(q, z=z'=0) + \frac{1}{2} \tilde{\varphi}_{\mathrm{scr}}(q, z=z'=d) - \tilde{\varphi}_{\mathrm{scr}}(q, z=d, z'=0) \Bigg) =\\
 \frac{e^2 N}{2 \pi} \lim_{q \rightarrow 0} \Bigg(\frac{2 \pi \left(1 - \exp{\left(-\frac{q d}{l}\right)}\right)}{q l} - \frac{4 \pi^2 e^2 \Pi(q)}{2 q^2 \left(1 + e^2 l^2 \frac{2 \pi}{q l} \Pi(q)\right)}\left(1 + \exp{\left(- 2 \frac{q d}{l}\right)} - 2 \exp{\left(- \frac{q d}{l}\right)}\right)\Bigg) =\\
  \frac{e^2 N}{2 \pi} \frac{2\pi d}{l^2} = \frac{e^2 N d}{l^2},
\end{multline}
where we have used the fact that $\Pi(q) = \Pi(q, \omega = 0) \sim \text{const}\times q^2 + O(q^4)$ for $q \rightarrow 0$.
\end{widetext}

One can see that the answer does not depend on the polarization function $\Pi(q)$ and coincides with the energy of a capacitor made of two parallel planes of area $S$ at a distance $d$ with charge $Q = e N$ on its plates. Indeed, such a capacitor has the capacitance $C = S/(4 \pi d)$ and energy $E_{\mathrm{cap}} = Q^2/(2 C)$. Since $S = 2 \pi l^2 N$, one easily finds that $E_{\mathrm{Hartree}} = E_{\mathrm{cap}}$.

So, in this simple model, the full Hartree energy of a filled Landau level and the compensating charge layer does not depend on screening. Of course, this model does not take into account many features of real experimental systems. However, it clearly illustrates that one can hardly expect to have different Hartree energies for differently screened potentials. Therefore, the Hartree energy does not contribute to the population reversal effect.

\section{\label{peculiDiag}Peculiarities of calculations of virtual hopping corrections}

In sub-subsection \ref{hoppingQM} we presented the derivation of the corrections to the pseudopotentials which correspond to the diagram in Fig.~\ref{fig:diag}b. The derivation presented there is based on a consideration of two interacting particles, which interact through a potential, within the methodology of first quantized quantum mechanics. In fact, the full expression for this correction comes from the consideration within a quantum field theory methodology and second quantization approach. Thus, the diagram should take into account the potential screening (including the polarization function frequency dispersion) and corrections to the electron propagator (self-energy). In this appendix we argue that the effects of interaction potential screening and the electron free energy are of the same order as the next perturbation theory correction.

First, we neglect corrections to the electron propagator. We denote the free electron propagator as $G_0(\omega, \mathbf{q}_1, \mathbf{q}_2, j)$ ($j$ is the LL in which the electron propagates; $\mathbf{q}_{1/2}$ are the 2-momenta~--- the electron propagator in uniform magnetic field is not translation-invariant, so it contains two momenta coming from the Fourier transforms with respect to two spatial coordinates). We denote the photon propagator as $D(\omega, \mathbf{q})$ (it is translation-invariant, so depends only on one spatial momentum). Then the propagators can be expressed as follows (infinitely small imaginary parts of the denominators are not written out explicitly):
\begin{equation}
G_0(\omega, \mathbf{q}_1, \mathbf{q}_2, j) = \frac{f_{j}(\mathbf{q}_1, \mathbf{q}_2)}{\omega - E_j},
\end{equation}
\begin{equation}
D(\omega, \mathbf{q}) = \frac{v(\mathbf{q})}{1 + l^2 v(\mathbf{q}) \Pi(\mathbf{q}, \omega)},
\end{equation}
$f_{j}(\mathbf{q}_1, \mathbf{q}_2)$ are smooth functions without poles, $\Pi(\mathbf{q}, \omega)$ is the polarization function, $v(\mathbf{q})$ is the unscreened Coulomb interaction potential, the product $v(\mathbf{q}) \Pi(\mathbf{q}, \omega)$ as a function of $\mathbf{q}$ is a smooth function without poles.

The diagram we are interested in can be expressed then via the propagators (we will need only the $\omega$ dependence for our analysis, so the dependence on the spatial momenta is not written explicitly):
\begin{multline}
\mathrm{Diagram} = \int_{\text{over momenta}} \int d\omega\ G(\omega,j_1) \times{}\\
{}D(E_1-\omega)G(-\omega+E_1+E_2,j_2)D(-\omega+E_1+E_2-E_4),
\end{multline}
where $E_i$ are the energies of the incoming and outgoing electrons. In our case both are in the same LL $j$: $E_i = E_{j}^{kin}$. $j_{1}$ and $j_{2}$ denote the levels to which the electrons hop, just like in sub-subsection~\ref{hoppingQM}.

All the functions (propagators) participating in the expression are the time-ordered correlation functions at zero temperature and their poles are situated in such a way that one can do Wick rotation $\omega \rightarrow i \omega$ (the zero of energies is the partially filled LL $E_{j}^{kin} = 0$). Then, concentrating on the denominators, we can write the diagram via the integral over Euclidean frequency:
\begin{multline}
\mathrm{Diagram} \propto \int_{\text{over momenta}} \int_{-\infty}^{\infty} d\omega \frac{...}{i \omega-E_{j_1}^{\mathrm{kin}}} \times{}\\
{} \frac{...}{-i\omega-E_{j_2}^{\mathrm{kin}}} \times
 \frac{...}{1 + l^2 v(...) \Pi(..., -i\omega)} \times{}\\
  {}\frac{...}{1 + l^2 v(...) \Pi(..., -i\omega)}.
\end{multline}
Evidently, the denominators of the electron propagators determine the region of frequencies which give dominating contribution to the integral. And this region is the interval centered at zero frequency with the width of the order of the energy distance to the closest higher LL.

Consider the difference of the full $D(\omega, \mathbf{q})$ and the bare $D_0(\omega, \mathbf{q}) = v(\mathbf{q})$ photon propagators:
\nmq{
D_0(\omega, \mathbf{q}) - D(\omega, \mathbf{q}) = \frac{l^2 v(\mathbf{q})^{2} \Pi(\mathbf{q}, \omega)}{1 + l^2 v(\mathbf{q}) \Pi(\mathbf{q}, \omega)} =\\ D(\omega, \mathbf{q}) \times O\left(\frac{e^2}{l \kappa \hbar \omega_c}\right).
}
Therefore, replacing $D(\omega, \mathbf{q})$ with $D_0(\omega, \mathbf{q})$ brings in a relative error proportional to the ratio between the typical interaction energy and the distance to the closest LL. Since we neglect next order perturbation theory correction, taking the screening into account would exceed the accuracy of our calculation. In principle, it is possible to justify use of the screened interaction by expanding in other parameters (e.g. 1/N expansion \cite{Aleiner2012}). However, in that case one needs to use not just the zero-frequency polarization function as in \eqref{screening}, but to take into account the polarization function frequency dispersion. We do not do this and use the unscreened interaction for the calculation.

Now we discuss the effect of corrections to the electron propagator on the diagram. These corrections should lead to LLs shifting and broadening. It is natural to expect that the shifts and level widths are of the order of the Coulomb interaction energy scale $e^2/(\kappa l)$. This would change the region of important frequencies by something proportional to the $e^2/(\kappa l)$.

Therefore, neglecting the effects of the polarization function and the electron free energy in the diagram introduces a relative error proportional to the ratio between the typical interaction energy and the distance to the closest LL. Since we neglect the next order perturbation theory correction (and, moreover, the second order three-body interaction), taking these effects into account would exceed the accuracy of our calculation.

\section{\label{PfaffSupplMat}Additional data on the Moore-Read Pfaffian state stability}

In this appendix we provide additional details regarding the stability of the Moore-Read state.

\begin{figure}[tbp]
  \centering
  \includegraphics[width=.96\columnwidth]{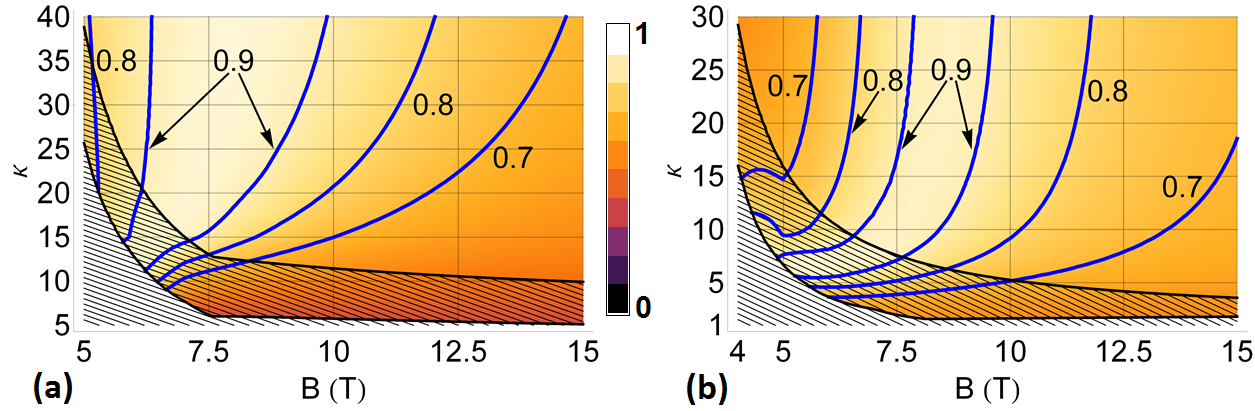}
  \caption{\textbf{Color plot of the overlap of the ground state with the Moore-Read Pfaffian for $12$ particles as a function of the magnetic field $B$ and the dielectric constant $\kappa$ with no virtual hopping corrections taken into account.} $(a)$~-- for the $(1, -1)$ LL at $U = 50\ \text{meV}$, $(b)$ -- for the $(2, -1)$ LL at $U = 30\ \text{meV}$. Contours show the lines of constant overlap. The region where perturbative analysis is not applicable according to the type C estimate is hatched. Data is not shown beyond the region where perturbative analysis is applicable according to the type S estimate. (Color online).}
  \label{fig:overlapNonHopp}
\end{figure}

Figures \ref{fig:overlapNonHopp}a and \ref{fig:overlapNonHopp}b show the dependence of the overlap of the exact ground state of the system with the Moore-Read Pfaffian for $N = 12$ particles on the magnetic field and the dielectric constant at $U = 50\ \text{meV}$ for the $(1, -1)$ LL and at $U = 30\ \text{meV}$ for the $(2, -1)$ LL respectively \textit{with no virtual hopping to the nearby LLs taken into account}. Screening is still taken into account. We do not show the data in the region where the perturbative analysis is not applicable according to type S estimate (with typical interaction energy scale needed for the estimate taken to be the \textit{screened} potential zeroth pseudopotential at the LL under consideration). The region of inapplicability of our theory according to the type C estimate (with typical interaction energy scale taken to be the bare \textit{Coulomb} potential zeroth pseudopotential) is hatched. As one can see (compare with Fig.~\ref{fig:overlap}), the effect virtual hopping corrections, defined in sub-subsection~\ref{hoppingQM}, have on the phase diagram within the region of applicability of perturbative treatment of LL mixing is very significant, especially for the $(1, -1)$ LL.

\begin{figure}[tbp]
  \centering
  \includegraphics[width=.96\columnwidth]{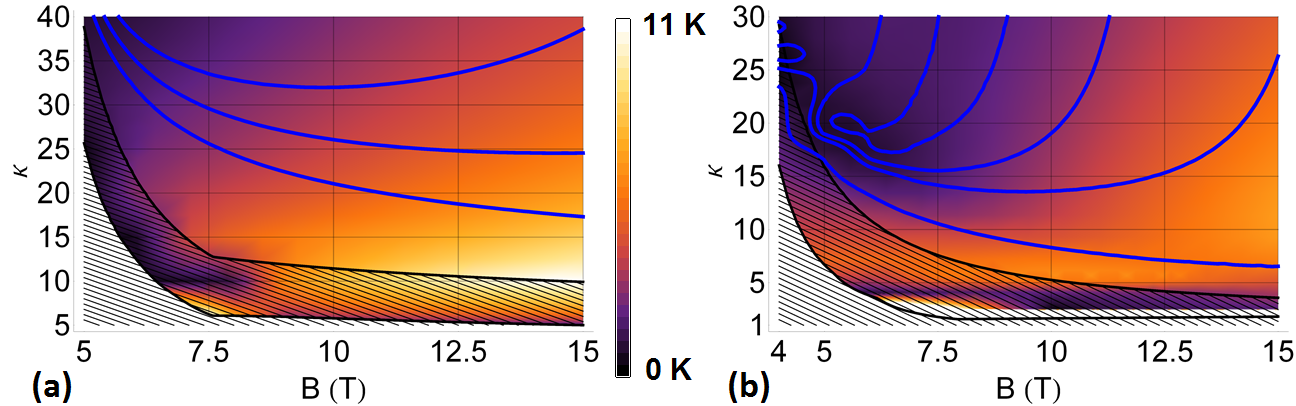}
  \caption{\textbf{Color plot of the gap between the ground state and the first excited state computed for $12$ particles as a function of the magnetic field $B$ and the dielectric constant $\kappa$.} $(a)$~-- for the $(1, -1)$ LL at $U = 50\ \text{meV}$, $(b)$ -- for the $(2, -1)$ LL at $U = 30\ \text{meV}$. The region where perturbative analysis is not applicable according to the type C estimate is hatched. Data is not shown beyond the region where perturbative analysis is applicable according to the type S estimate. The blue lines coincide with the overlap level lines from Fig.~\ref{fig:overlap}. (Color online).}
  \label{fig:gap}
\end{figure}

\begin{figure}[tbp]
  \centering
  \includegraphics[width=.96\columnwidth]{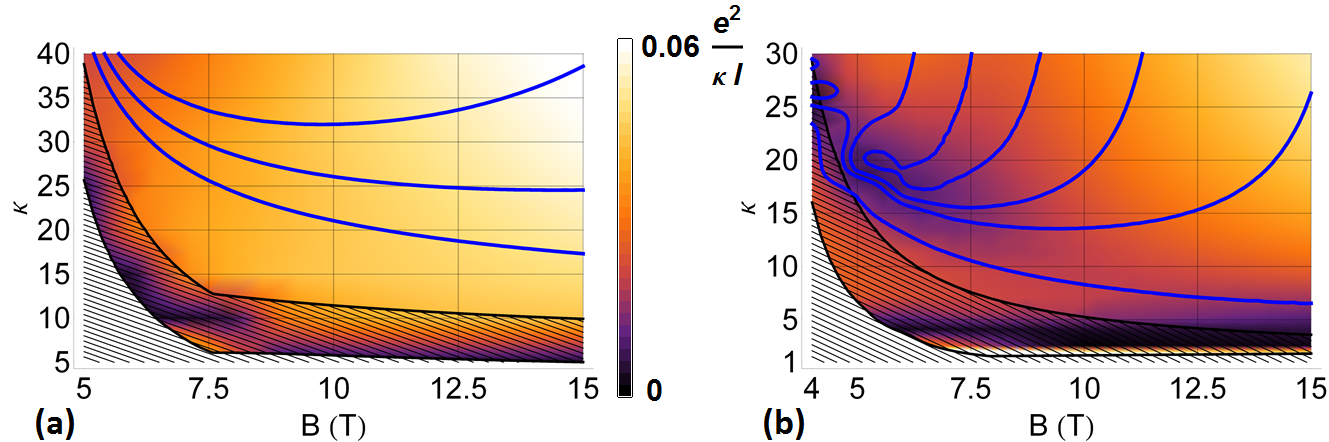}
  \caption{\textbf{Color plot of the gap between the ground state and the first excited state computed for $12$ particles as a function of the magnetic field $B$ and the dielectric constant $\kappa$.} $(a)$~-- for the $(1, -1)$ LL at $U = 50\ \text{meV}$, $(b)$ -- for the $(2, -1)$ LL at $U = 30\ \text{meV}$. The region where perturbative analysis is not applicable according to the type C estimate is hatched. Data is not shown beyond the region where perturbative analysis is applicable according to the type S estimate. The blue lines coincide with the overlap level lines from Fig.~\ref{fig:overlap}. (Color online).}
  \label{fig:gapCoul}
\end{figure}

Figures \ref{fig:gap}a and \ref{fig:gap}b show the numerically found gap between the exact ground state of the system with $12$ particles and its exact first excited state. In these figures, the gap is plotted as a function of the magnetic field and the dielectric constant at $U = 50\ \text{meV}$ for the $(1, -1)$ LL (Fig.~\ref{fig:gap}a) and at $U = 30\ \text{meV}$ for the $(2, -1)$ LL (Fig.~\ref{fig:gap}b) respectively. Virtual hopping to the nearby LL \textit{is} taken into account. We do not show the data in the region where the perturbative analysis is not applicable according to type S estimate. The region of inapplicability of our theory according to the type C estimate is hatched. The blue lines are the overlap level lines from Fig.~\ref{fig:overlap}. Figures \ref{fig:gapCoul}a and \ref{fig:gapCoul}b show the same data on the gaps in the units of typical Coulomb energy $e^2/(\kappa l)$, where $l$ is the magnetic length $l = \sqrt{\hbar c/(e B)}$.

The behaviour of the gaps for the two levels is quite different. For both levels the gaps generally increase as the magnetic field increases and as the dielectric constant decreases. This can be partially (for magnetic field) or fully (for the dielectric constant) attributed to the increase of $e^2/\kappa l$. However, there are some interesting features in the plots. For the $(1, -1)$ LL the gap drops to almost zero at a line in the hatched region. For the $(2, -1)$ LL there are two such features: one in the hatched region and another slightly above the hatched region (better seen in Fig.~\ref{fig:gapCoul}b than in Fig.~\ref{fig:gap}b). Although two of these features are situated in the hatched region where the applicability of our methodology can be questioned, and the third one is much less pronounced, these feature, obtained for a finite number of particles, may suggest that a transition between different topological orders happens across those lines. Such transitions are likely to happen at small dielectric constants where the system physics becomes essentially non-single Landau level.

We remind the reader that for the $(2, -1)$ LL the high overlap with the Moore-Read Pfaffian state is achieved near $B = 8 \mathrm{T}$ at $\kappa \gtrsim 20$, as can be seen from Fig.~\ref{fig:overlap}.

\begin{figure}[tbp]
  \centering
  \includegraphics[width=.96\columnwidth]{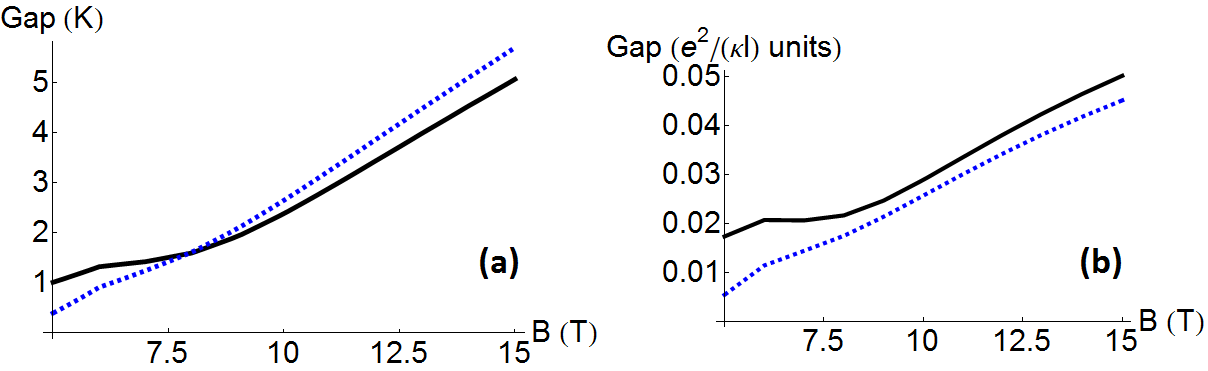}
  \caption{\textbf{Dependence of the gap between the ground state and the first excited state computed for $12$ particles at the $(2, -1)$ LL for $U = 30\ \text{meV}$ and $\kappa = 20$ (blue dashed line) and $25$ (black solid line) as a function of the magnetic field $B$.} $(a)$ -- in $K$, $(b)$ -- in $e^2/(\kappa l)$ units. Only the part where perturbative analysis is applicable according to the type S estimate is shown. (Color online).}
  \label{fig:gap2fixedkappa}
\end{figure}

Figures \ref{fig:gap2fixedkappa}a and \ref{fig:gap2fixedkappa}b show the dependence of the gap on the magnetic field at $U = 30\ \text{meV}$ and $\kappa = 20\text{ and }25$ for the $(2, -1)$ LL in $K$ and in $e^2/(\kappa l)$ units respectively.

\begin{figure}[tbp]
  \centering
  \includegraphics[width=.96\columnwidth]{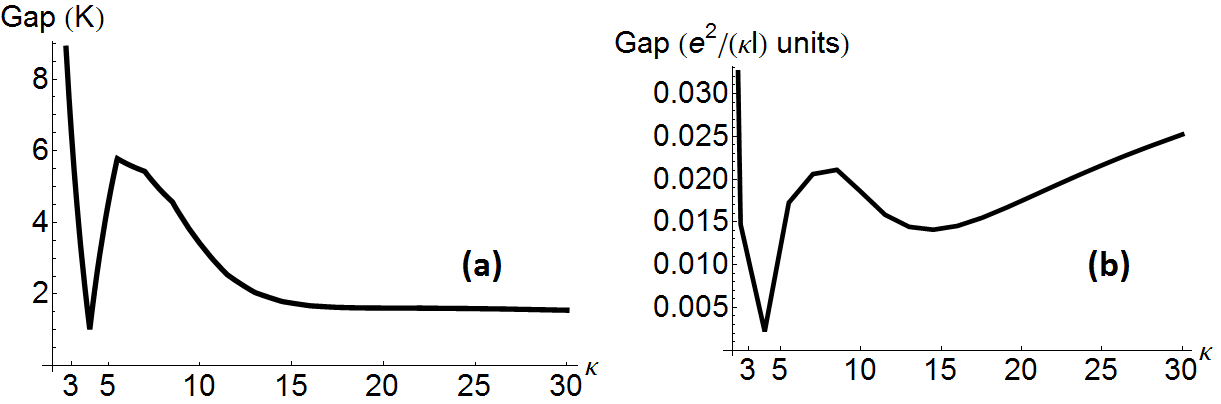}
  \caption{\textbf{Dependence of the gap between the ground state and the first excited state computed for $12$ particles at the $(2, -1)$ LL for $U = 30\ \text{meV}$ and $B = 8\ \mathrm{T}$ as a function of the dielectric constant $\kappa$.} $(a)$ -- in $K$, $(b)$ -- in $e^2/(\kappa l)$ units. Only the part where perturbative analysis is applicable according to the type S estimate is shown. The cusps present in the plots are due to the fact that simulations were performed at a discrete set of dielectric constants $\kappa$ with a step $\delta \kappa = 1.5$. Therefore, the data accurately represents the region where the dependence is smooth, while where the gap changes quickly the true data between the cusp points may deviate from the lines drawn.}
  \label{fig:gap2fixedB8}
\end{figure}

Figures \ref{fig:gap2fixedB8}a and \ref{fig:gap2fixedB8}b show the dependence of the gap on the dielectric constant at $U = 30\ \text{meV}$ and fixed $B = 8\ \mathrm{T}$ for the $(2, -1)$ LL in $K$ and in $e^2/(\kappa l)$ units respectively.

Now we give some data on overlaps and gaps from numerical diagonalization for different numbers of particles $N = 8,10,12,14$.

\begin{figure}[tbp]
  \centering
  \includegraphics[width=.96\columnwidth]{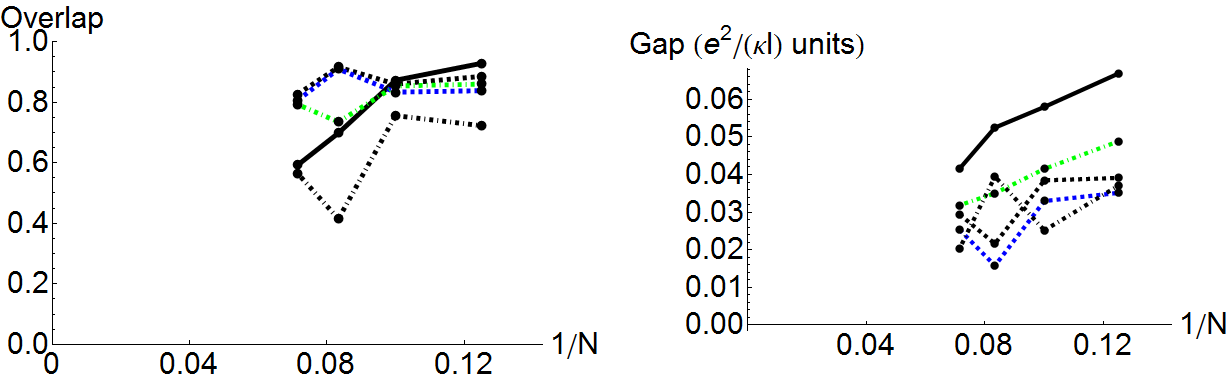}
  \caption{\textbf{Dependence of the overlap with the Pfaffian and gap to the first excited state on the number of particles $N$ ($N = 8,10,12,14$).} Solid black line is for the non-relativistic $n = 1$ LL. Dashed lines are for the $(2, -1)$ LL at $U = 30\ \text{meV}$: at $\kappa = 25$ and $B = 8\ \text{T}$ (black), at $\kappa = 20$ and $B = 7.5\ \text{T}$ (blue). Dot-dashed lines are for the $(1, -1)$ LL at $U = 50\ \text{meV}$: at $\kappa = 25$ and $B = 8\ \text{T}$ (black), at $\kappa = 40$ and $B = 8\ \text{T}$ (green). The gap is presented in units of typical Coulomb energy $e^2/(\kappa l)$. (Color online).}
  \label{fig:diffnumpart}
\end{figure}

Figure \ref{fig:diffnumpart} compares the data on the overlap with the Moore-Read Pfaffian and the gap to the first excited state for the $(1, -1)$ and the $(2, -1)$ LLs in BLG and the non-relativistic $n = 1$ LL. The data are shown for different numbers of particles $N = 8,10,12,14$. For the $(1, -1)$ and the $(2, -1)$ LLs in BLG external parameters are set to be near the maximum overlap regions (according to Fig.~\ref{fig:overlap}) at several values of the dielectric constant $\kappa$; we also present the data for $B = 8$, $\kappa = 25$ for the $(1, -1)$ LL. The gap is presented in units of the typical Coulomb energy $e^2/(\kappa l)$. The overlaps for the $(2, -1)$ LL, as well as for the $(1, -1)$ LL at $\kappa = 40$, appear to be more stable as the number of particles grows than for the non-relativistic system. It is interesting to note that for the $(2, -1)$ LL the overlap is slightly higher at $N = 12$ particles than at $N = 8, 10,\text{ and }14$, while for the $(1, -1)$ LL it is vice versa.

\begin{figure}[tbp]
  \centering
  \includegraphics[width=.96\columnwidth]{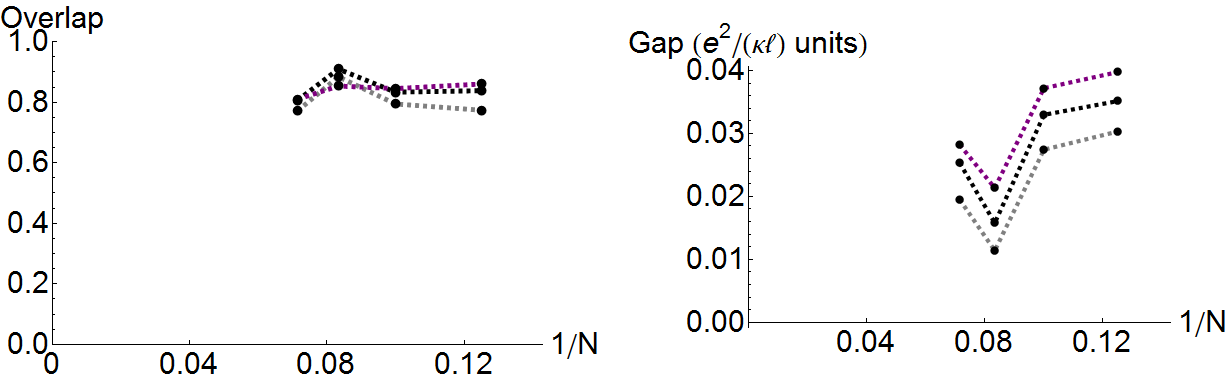}
  \caption{\textbf{Dependence of the overlap with the Moore-Read Pfaffian and gap to the first excited state on the number of particles $N$ ($N = 8,10,12,14$) for the $(2, -1)$ LL at $U = 30\ \text{meV}$, $\kappa = 20$.} Shown are the dependences for $B = 7.5\ \text{T}$ (dashed black line), $B = 6\ \text{T}$ (dashed grey line), and $B = 9\ \text{T}$ (dashed purple line). (Color online).}
  \label{fig:diffnumpartN2eps20}
\end{figure}

\begin{figure}[tbp]
  \centering
  \includegraphics[width=.96\columnwidth]{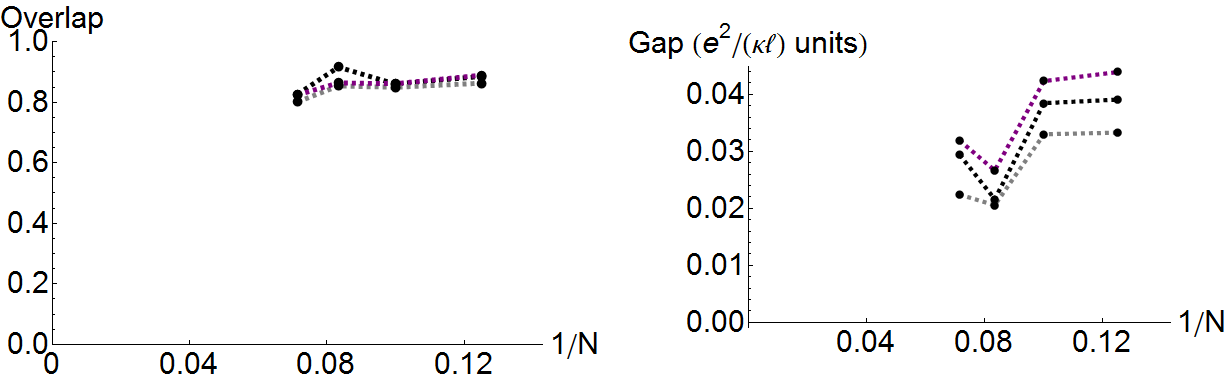}
  \caption{\textbf{Dependence of the overlap with the Moore-Read Pfaffian and gap to the first excited state on the number of particles $N$ ($N = 8,10,12,14$) for the $(2, -1)$ LL at $U = 30\ \text{meV}$, $\kappa = 25$.} Shown are the dependences for $B = 8\ \text{T}$ (dashed black line), $B = 6.5\ \text{T}$ (dashed grey line), and $B = 9.5\ \text{T}$ (dashed purple line). (Color online).}
  \label{fig:diffnumpartN2eps25}
\end{figure}

Figures \ref{fig:diffnumpartN2eps20}, \ref{fig:diffnumpartN2eps25} present the dependence of the overlap and the gap at several parameter points for the $(2, -1)$ LL. Points are taken to be at the same mini-gap and the dielectric constant values as in the previous figure, but the magnetic field changes. We took the points in the maximum overlap region, slightly to the left and to the right of it.

\clearpage

\bibliography{Thesis}
\end{document}